\documentclass[a4paper,11pt]{article}
\pdfoutput=1 

\usepackage{jcappub} 

\usepackage[T1]{fontenc} 

\usepackage{hyperref}
\usepackage{graphicx}
\usepackage{url}
\usepackage{amsmath}
\usepackage{amsfonts}
\usepackage{amssymb}

\def\beq{\begin{equation}}
\def\eeq{\end{equation}}
\def\beqa{\begin{eqnarray}}
\def\eeqa{\end{eqnarray}}
\newcommand{\ts}{\thinspace}

\title{Cosmological Constraints on the  Gravitational Interactions of Matter and Dark Matter
}

\author[a]{Yang Bai,}
\author[a,b]{Jordi Salvado,}
\author[a]{Ben A. Stefanek}

\affiliation[a]{Department of Physics, University of Wisconsin-Madison, Madison, WI 53706, USA}
\affiliation[b]{Wisconsin IceCube Particle Astrophysics Center, Madison, WI 53706, USA}

\emailAdd{yangbai@physics.wisc.edu}
\emailAdd{salvadoserra@wisc.edu}
\emailAdd{stefanek@wisc.edu}

\abstract{Although there is overwhelming evidence of dark matter from its gravitational interaction, we still do not know its precise gravitational interaction strength or whether it obeys the equivalence principle. Using the latest available cosmological data and working within the framework of $\Lambda\mbox{CDM}$, we first update the measurement of the Newton's constant for all matter: $G_N=7.26^{+0.27}_{-0.27}\times 10^{-11}\,\mbox{m}^{3}\mbox{kg}^{-1}\mbox{s}^{-2}$, which differs by $2.2 \sigma$ from the standard laboratory-based value. In general relativity, dark matter equivalence principle breaking can be mimicked by a long-range dark matter force mediated by an ultra light scalar field. Using the Planck three year data, we find that the dark matter ``fifth-force'' strength is constrained to be weaker than $10^{-4}$ of the gravitational force. We also introduce a phenomenological, post-Newtonian two-fluid description to explicitly break the equivalence principle by introducing a difference between dark matter inertial and gravitational masses. Depending on the decoupling time of the dark matter and ordinary matter fluids, the ratio of the dark matter gravitational mass to inertial mass is constrained to be unity at the $10^{-6}$ level.
}

\begin{document}
\maketitle
\flushbottom

\section{Introduction}
\label{sec:intro}
Although Newton's description of gravity has existed for over three hundred years, the value of the Newtonian gravitational constant $G_N$ has only been measured at the $10^{-4}$ level. This is in contrast to the fine-structure constant, which has been measured very accurately with a relative error of $10^{-9}$. Over the past thirty years, there has been a large effort to precisely measure the value of the Newtonian gravitational constant in the laboratory. Most of these measurements were torsion balance based and achieved a level of accuracy which can constrain many new physics scenarios (see Ref.~\cite{Adelberger:2003zx} for a review). However, these laboratory based measurements suffer from systematic errors and have results which are inconsistent with each other. As a result, the discrepancy between the measurements lead the Committee on Data for Science and Technology (CODATA), which determines the internationally accepted standard values, to increase the relative uncertainty of the 2010 recommended value of $G_{N}$ by 20\% to 120 parts per million compared to that of the 2006 value~\cite{Mohr:2012tt}. Additionally, a recent measurement using cold atom interferometry obtains similar accuracy~\cite{cold-atom}, but the central value is in tension with the CODATA standard value at 1.5$\sigma$.

In addition to laboratory-based measurements of $G_N$, cosmological data can also provide an independent measurement at a much larger distance scale and also serve as a consistency check. As pointed out in Ref.~\cite{Zahn:2002rr} and employed in Refs.~\cite{Umezu:2005ee,Galli:2009pr}, the measurements of the Cosmic Microwave Background (CMB) temperature anisotropy and polarization can be used to extract the value of $G_{N}$ from cosmology. In the last few years, an enormous amount of cosmological data both from satellite and ground-based experiments has been gathered. The Planck satellite in particular has ushered in the era of precision cosmology, measuring the CMB temperature and polarization anisotropy with unparalleled accuracy~\cite{Ade:2013zuv,Ade:2015xua}. In addition, ground-based telescopes such as the Atacama Cosmology Telescope (ACT)~\cite{Sievers:2013ica} and the South Pole Telescope (SPT)~\cite{Hou:2012xq}, provide high accuracy polarization measurements at high multipole moments. In this work, we will update the cosmological measurement of $G_{N}$ for all matter using the latest available cosmological data, including the 2013 Planck release, ACT/SPT, and Big Bang Nucleosynthesis (BBN).

Another interesting question that may be answered using cosmological data is whether dark matter has the same gravitational interaction as the ordinary baryonic matter. So far we have seen extensive evidence for the existence of dark matter, such as galactic rotation curves and gravitational lensing. There is no doubt that dark matter has gravitational interactions, although we have not found additional forces felt by dark matter, despite a large effort to understand dark matter particle properties. This serves as motivation to understand the dark matter gravitational interaction better. In this paper, we will use the current cosmological data to constrain the properties of the dark matter gravitational interaction.

Different gravitational interactions for different matter types inevitably break the Weak Equivalence Principle (WEP), which states that all objects in a uniform gravitational field, independent of the mass or other compositional properties, will experience the same acceleration. In Newtonian language, the difference between inertial mass (the mass appearing in Newton's second law) and gravitational mass (the mass appearing in Newton's law of gravity) must be exactly zero for the WEP to be respected. For ordinary baryonic matter, modern experiments using torsion balances report that the difference between inertial and gravitational masses is zero at the $10^{-13}$ level~\cite{Wagner:2012}. Thus, violations of the WEP in the visible sector are tightly constrained. The constraint on WEP breaking in the dark sector is much less restrictive. 

One simple way to mimic dark matter equivalence principle breaking is to introduce a long-range force only for dark matter~\cite{Friedman:1991dj,Bean:2001ys,Gubser:2004uh,Nusser:2004qu,Bean:2008ac}. In this class of models, the gravitational interaction strengths for dark matter and baryonic matter are the same, although the total long range force between two clumps of dark matter is different from baryonic matter. In our study, we will also update the cosmological constraints on the fifth force strength and pay additional attention to the force carrier energy density. 

Beyond the fifth-force description, we also introduce a phenomenological, post-Newtonian model to explicity distinguish dark matter gravitational and inertial mass. For ordinary cosmology, one can derive parts of the background Friedmann and linear perturbation equations in the post-Newtonian description by treating the scale factor as an expanding sphere filled with a homogeneous and isotropic fluid~\cite{Mukhanov-book}. For pressureless matter, the continuity and Euler equations of fluid mechanics can be matched to the energy-momentum conservation equations in general relativity. We will follow this description and treat baryonic matter and dark matter as two separate but coupled fluids. Effectively, the dark matter fluid is evolving with a different scale factor such that the ratio of dark matter gravitational to inertial mass can not be simply absorbed by the dark matter energy density. The cosmological data will provide a constraint for this ratio. A relevant example from the literature uses the tidal disruption of the Sagittarius dwarf galaxy orbiting the Milky Way to constrain this ratio at the 10\% level ~\cite{Kesden:2006zb}. We will find that cosmological data can provide a much more stringent constraint. 

Our paper is organized as follows. We first constrain Newton's constant for all matter in section~\ref{sec:all-matter}. Then, in section~\ref{sec:fifth-force} we update the constraints on dark matter fifth forces and in section~\ref{sec:two-fluid} we develop a two-fluid description to constrain dark matter WEP breaking. Finally we conclude our paper in section~\ref{sec:conclusion}. 

\section{Newton's Constant for all Matter}
\label{sec:all-matter}
It is well-known that the gravitational or Newton force of a probing body only depends on the product of the Newton's Constant $G$ and the central body mass. To break this degeneracy, an additional force (such as the electromagnetic or the weak interaction) is required to define the central body mass. Because of this fact, the existing studies in the literature~\cite{Zahn:2002rr,Umezu:2005ee,Galli:2009pr,Galli:2010it} have used data from the primordial abundances of light elements synthesized by BBN and CMB anisotropies to constrain $G$, or equivalently other fundamental constants like the fine-structure constant~\cite{Ade:2014lua}. In this section, we use the currently available cosmological data to derive a constraint on Newton's constant.

Before we go into our detailed analysis, we introduce a dimensionless parameter $\lambda_{G}$ to quantify the potential deviation of Newton's constant $G$ from the value, $G_N$, measured in the laboratory-based experiments
\beqa
G = \lambda^{2}_{G}\,G_N \,.
\label{eq:lambda-def}
\eeqa
We will use the currently suggested central value of $G_N = 6.67384(80)\times 10^{-11}\,\mbox{m}^{3}\mbox{kg}^{-1}\mbox{s}^{-2}$ from CODATA 2010~\cite{Mohr:2012tt}.~\footnote{There are large inconsistencies among different measurements. The CODATA Task group has taken the 11 values after each of their uncertainties multiplied by an ad-hoc factor of 14.} The relative error is $1.2\times 10^{-4}$. The latest measurement using laser-cooled atoms and quantum interferometry reaches the similar precision and has an agreement with the CODATA value at $1.5\sigma$~\cite{cold-atom}. The constraints from cosmological data will serve as an independent, time-sensitive measurement at large length scales.

\subsection{Dependence of CMB Anisotropy on Newton's Gravitational Constant}
With the introduction of $\lambda_G$ in Eq.~(\ref{eq:lambda-def}), the Friedmann equation is modified as
\beqa
{\cal H}^{2} = \left( \frac{\dot{a} }{a} \right)^2 =   \frac{8\pi}{3}\,a^2\,\lambda^{2}_{G}\,G_N\,\rho \,,
\label{eq:MFE}
\eeqa
where ${\cal H}$ is the Hubble rate of expansion; $a$ is the scale factor of the universe; $\rho$ is the total energy density; dot indicates derivative with respect to the conformal time $\tau$ with $dt = a(\tau) d\tau$. From Eq.~(\ref{eq:MFE}) we see that the effect of $\lambda_{G}$ is to change the amplitude of the expansion rate ${\cal H}(\lambda_{G}, a)$ by a factor of $\lambda_{G}$, but not the shape. Since no new preferred cosmological scale is introduced by varying $G$, a variation in Newton's gravitational constant is equivalent to a simple rescaling of the wave numbers, as pointed out in Ref.~\cite{Zahn:2002rr}. Using the baryonic equations of motion in the conformal Newtonian gauge~\cite{Ma:1995ey}, one has
\beqa
\dot{\delta}_b &=& - \theta_b + 3\,\dot{\phi} \,, \label{eq:baryon-delta}\\
\dot{\theta}_b &=& -\frac{\dot{a}}{a} \theta_b + c_s^2\,k^2\,\delta_b + \frac{4 \bar{\rho}_\gamma}{3\bar{\rho}_b}\,a n_e\sigma_T ( \theta_\gamma - \theta_b) + k^2 \psi\,.   \label{eq:baryon-theta}
\eeqa
Here, $\delta_b \equiv \delta \rho_b /\bar{\rho}_b$; $\theta_b \equiv i k_j v^j_b$; $c_s$ is the baryon speed of sound; $n_e$ is the free electron number density; $\sigma_T = 0.6652\times 10^{-24}$~cm$^2$ is the Thomson scattering cross section; $\phi$ and $\psi$ are the scalar metric perturbation. If $\sigma_T$ was zero or there was no Coulomb interaction, one can show that Eqns.~(\ref{eq:MFE})(\ref{eq:baryon-delta})(\ref{eq:baryon-theta}) are independent of $\lambda_G$ by a replacement of $\tau \rightarrow  \tau  \lambda_G$ and $k \rightarrow  k/\lambda_G$. The density fluctuations produced by a mode of wave-vector $k$ in a universe with $\lambda_{G} \neq 1$ have dynamics equivalent to a mode with $k' = k/\lambda_{G}$ in a universe with $\lambda_{G} = 1$. Thus, for primordial fluctuations governed by a power law, changing $G$ can be compensated by adjusting the amplitude of the primordial power spectrum appropriately. As a result, the CMB anisotropy spectrum does not change when varying $\lambda_{G}$ if there is no Coulomb interaction. This makes sense physically because the effect of changing $G$ is to cause the universe to expand slightly faster or slower by a factor of $\lambda_{G}$, causing the ``expansion clock'' to run at a different rate. Since we only measure angles in the CMB through ratios of distances, such a change precisely cancels.

To observe the effect of different values of $G$ in the CMB, we need an independent clock that measures the expansion rate. This independent clock will come from the physics of recombination through an additional Coulomb force acting on the baryons. The number density, $n_{e}$, of free electrons will then depend on $\lambda_{G}$. As we will show, this change in $n_{e}$ is observable in the CMB through its effect on the visibility function that enters into the integral solution for the temperature anisotropies produced by a mode of wave-vector \textbf{k} observed towards direction $\bf{\hat{n}}$~\cite{Seljak:1996is}. The temperature at a direction $\bf{\hat{n}}$ is an integration of the sources in the line of sight convoluted by the visibility function $g(\tau)$, which is related to the free electron number density $n_e(\tau)$ through the Thomson scattering as
\beqa
g(\tau) = - \frac{d}{d\tau}\left[ \exp(-\kappa)  \right]\,, \qquad
\mbox{with} \quad 
\kappa(\tau) \equiv \sigma_{T} \int_{\tau}^{\tau_{0}} a(\tau)\, n_{e}(\tau) d\tau   \,.
\eeqa
Here, $\tau_0$ is the current universe time. The CMB anisotropy dependence on the visibility function is precisely what makes a change in the gravitational constant detectable. This is because the free electron density $n_{e}$ depends on the physics of recombination through the ionization fraction $x_{e} = n_{e}/n_{H}$ which evolves according to~\cite{Ma:1995ey}
\begin{equation}
\frac{dx_{e}}{d\tau} = a\,C_{r} \left[ \beta(T_{b})(1-x_{e}) - n_{H} \alpha^{(2)}(T_{b})\,x_{e}^{2} \right] \,,
\label{eq:dxdt}
\end{equation}
where $C_r$ is the Peebles correction factor to account the presence of non-thermal Lyman-$\alpha$ resonance photons; $T_b$ is the baryon temperature; $n_H$ is the number density of hydrogen atoms. Here, $\beta(T_{b})$ is the collisional ionization rate from the ground state
\beqa
\beta(T_{b}) = \left( \frac{m_{e}k_{B}T_{b}}{2\pi \hbar^{2}}\right)^{3/2} \ts e^{-B_{1}/k_{B}T_{b}}\alpha^{(2)}(T_{b})\,, 
\eeqa
with $B_{1} = 13.6 \ts \text{eV}$ and $\alpha^{(2)}(T_{b})$ as the recombination rate to excited states
\beqa
\alpha^{(2)}(T_{b}) = \frac{64\pi}{(27\pi)^{1/2}}\frac{e^{4}}{m_{e}^{2}c^{3}}\left(\frac{k_{B}T_{b}}{B_{1}}\right)^{-1/2} \ts \phi_{2}(T_{b}), \quad \phi_{2}(T_{b}) \approx 0.448 \ts \ln\left(\frac{B_{1}}{k_{B}T_{b}}\right)\, .
\eeqa

\begin{figure}[th!]
\begin{center}
\hspace*{-0.0cm}
\includegraphics[width=0.48\textwidth,clip=true,viewport= 20 0 590 360]{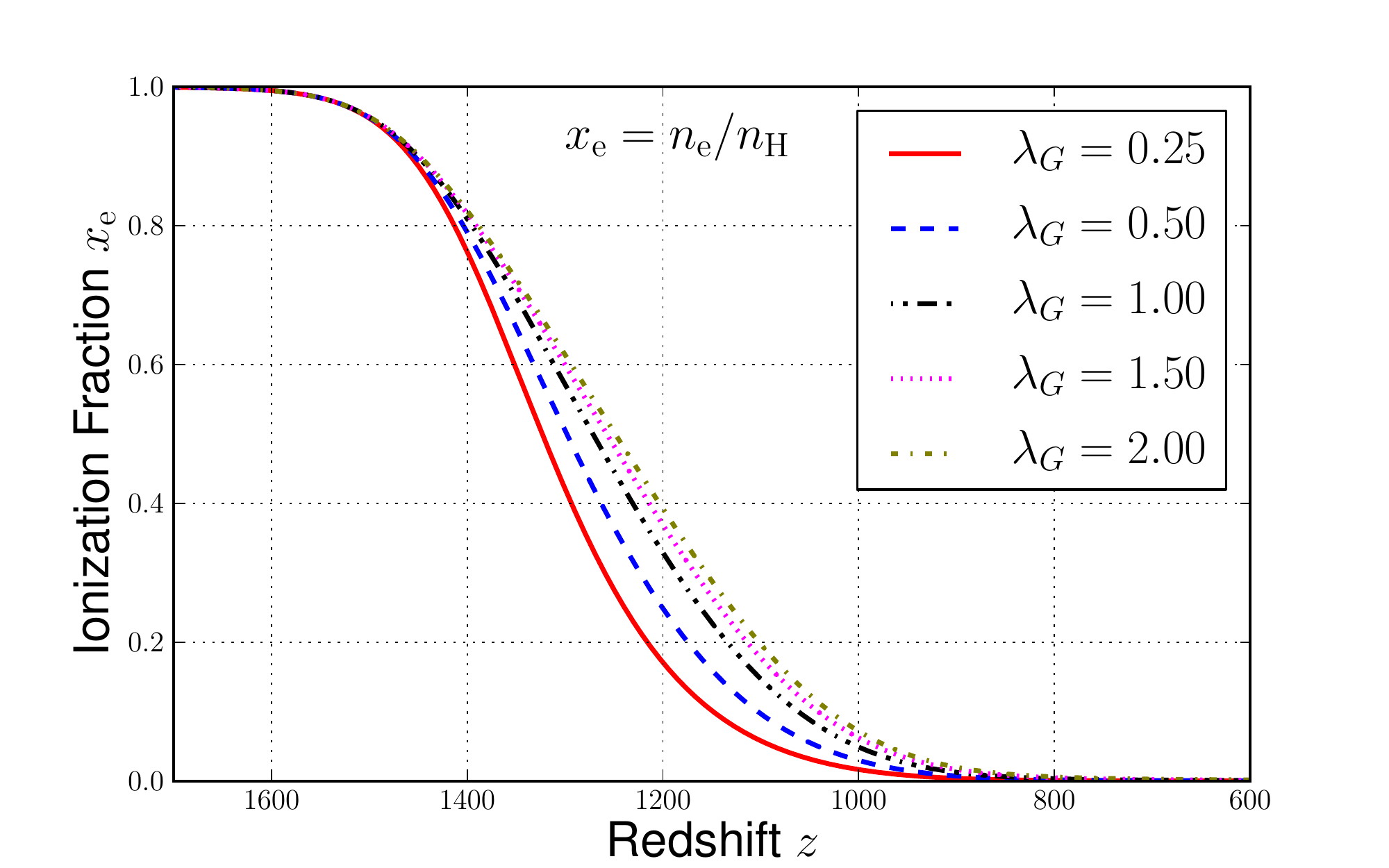} 
\hspace{0mm}
\includegraphics[width=0.48\textwidth,clip=true,viewport= 20 0 590 360]{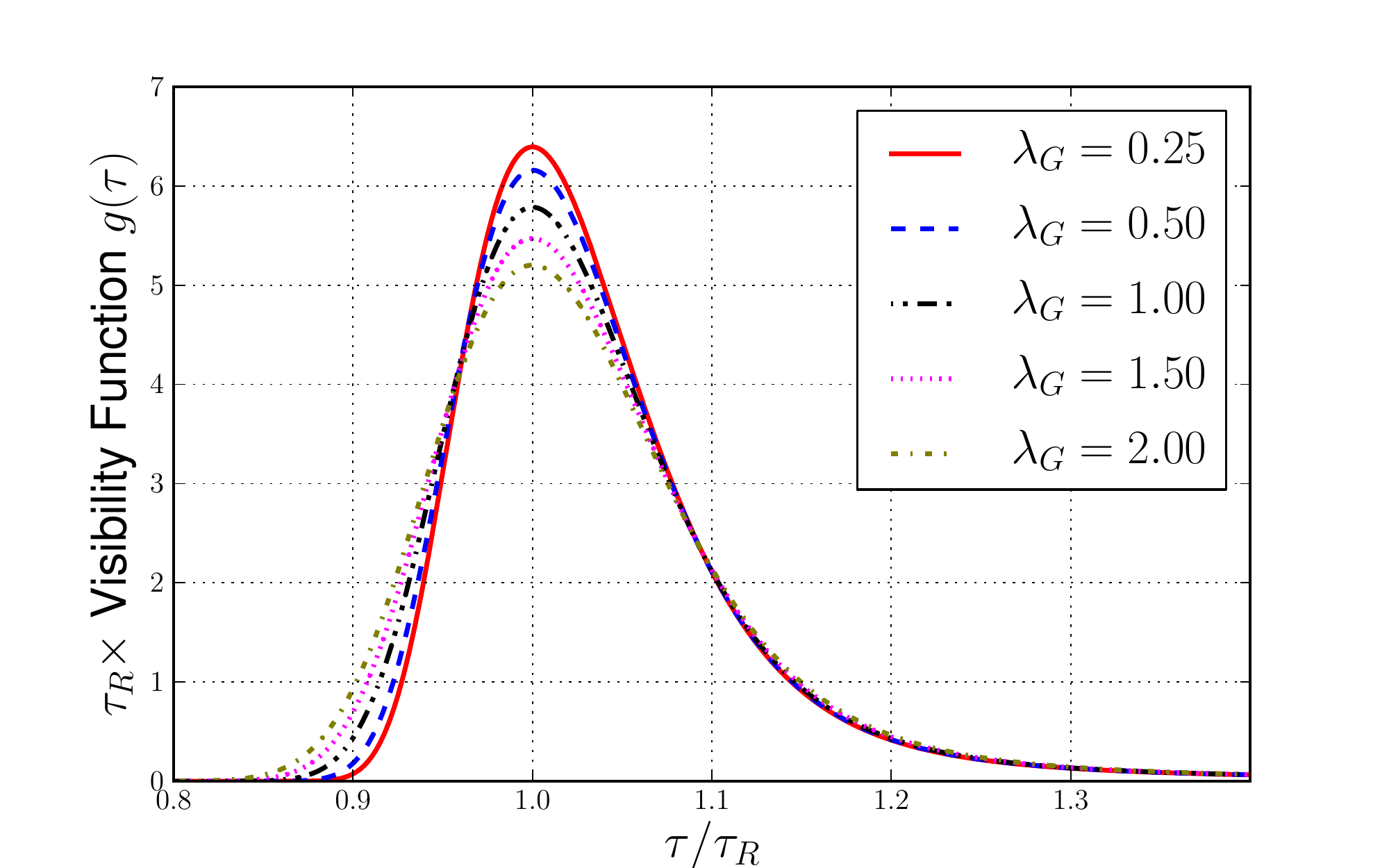}
\caption{{\bf Left panel:} the ionization fraction $x_e$ as a function of $z$ for different $\lambda_G$. {\bf Right panel:} the visibility function $g(\tau)$ as a function of conformal time. The axes have been rescaled using the recombination time $\tau_R$ to remove the overall scaling of $\tau$ with $\lambda_{G}$.}
\label{fig:Xe-g}
\end{center}
\end{figure}

Changing variables from $\tau$ to the scale factor $a$, $d/d\tau = a {\cal H} d/da = \lambda_G \, f(a) \,a\, d/da$, we have Eq.~(\ref{eq:dxdt}) in a different form
\begin{equation}
\frac{dx_{e}}{da} = \frac{C_{r}}{\lambda_{G}\,f(a)} \left[ \beta(T_{b})(1-x_{e}) - n_{H} \alpha^{(2)}(T_{b})x_{e}^{2} \right] \,.
\end{equation}
The above equation shows that the form  of $x_e(a)$ depends on $\lambda_G$. For a larger value of $\lambda_{G}$, this form shows a larger value of $x_e$ for a given $a$ or redshift. This can be understood as that for a larger $\lambda_{G}$, the universe expands faster at a given redshift, since it becomes more difficult  for hydrogen to recombine and this leads to a larger value for $x_{e}$. This effect is demonstrated in the left panel of Fig.~\ref{fig:Xe-g} in terms of the redshift $z = 1/a-1$.

\begin{figure}[th!]
\begin{center}
\hspace*{-0.0cm}
\includegraphics[width=0.49\textwidth,clip=true]{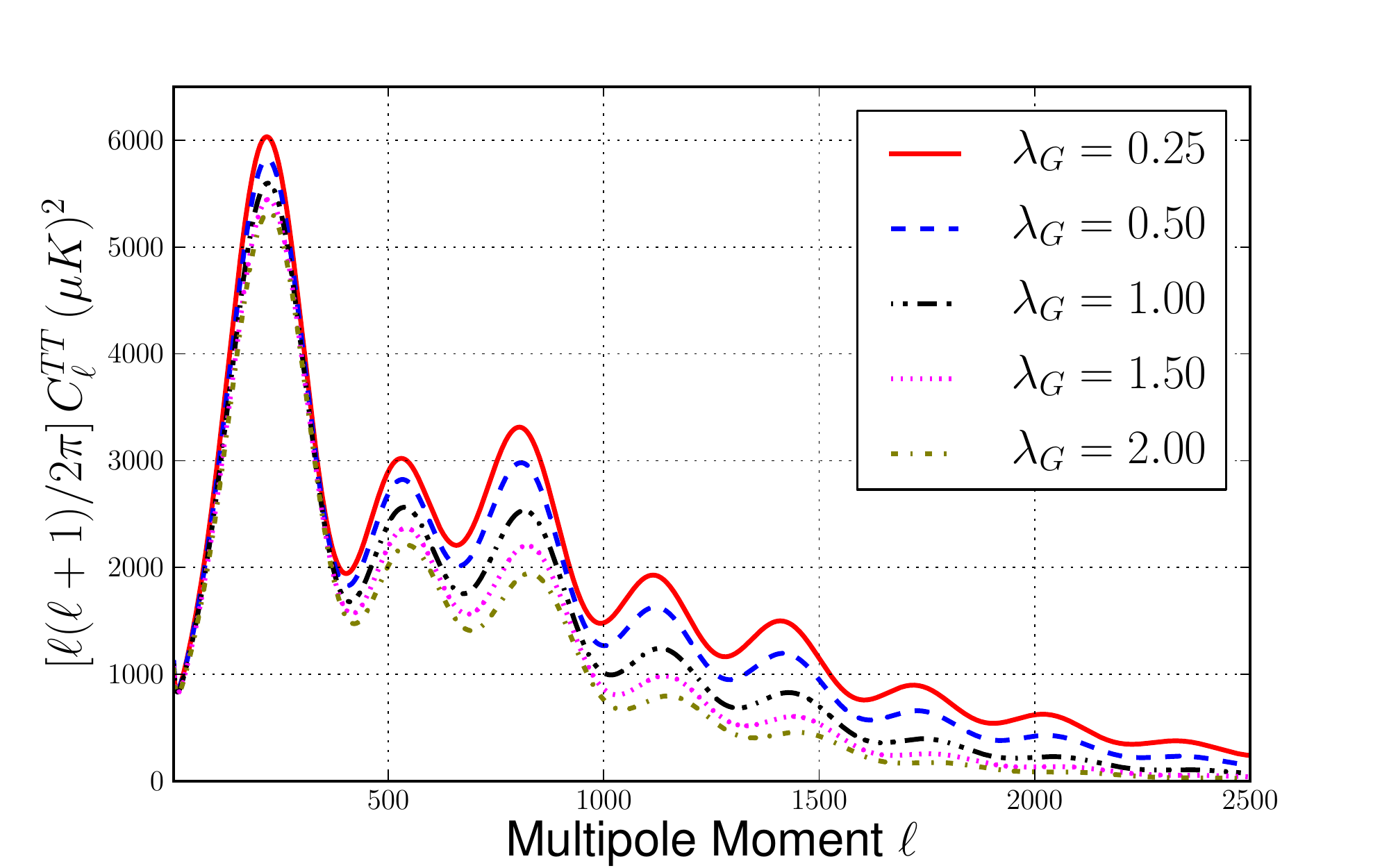}
\includegraphics[width=0.49\textwidth,clip=true]{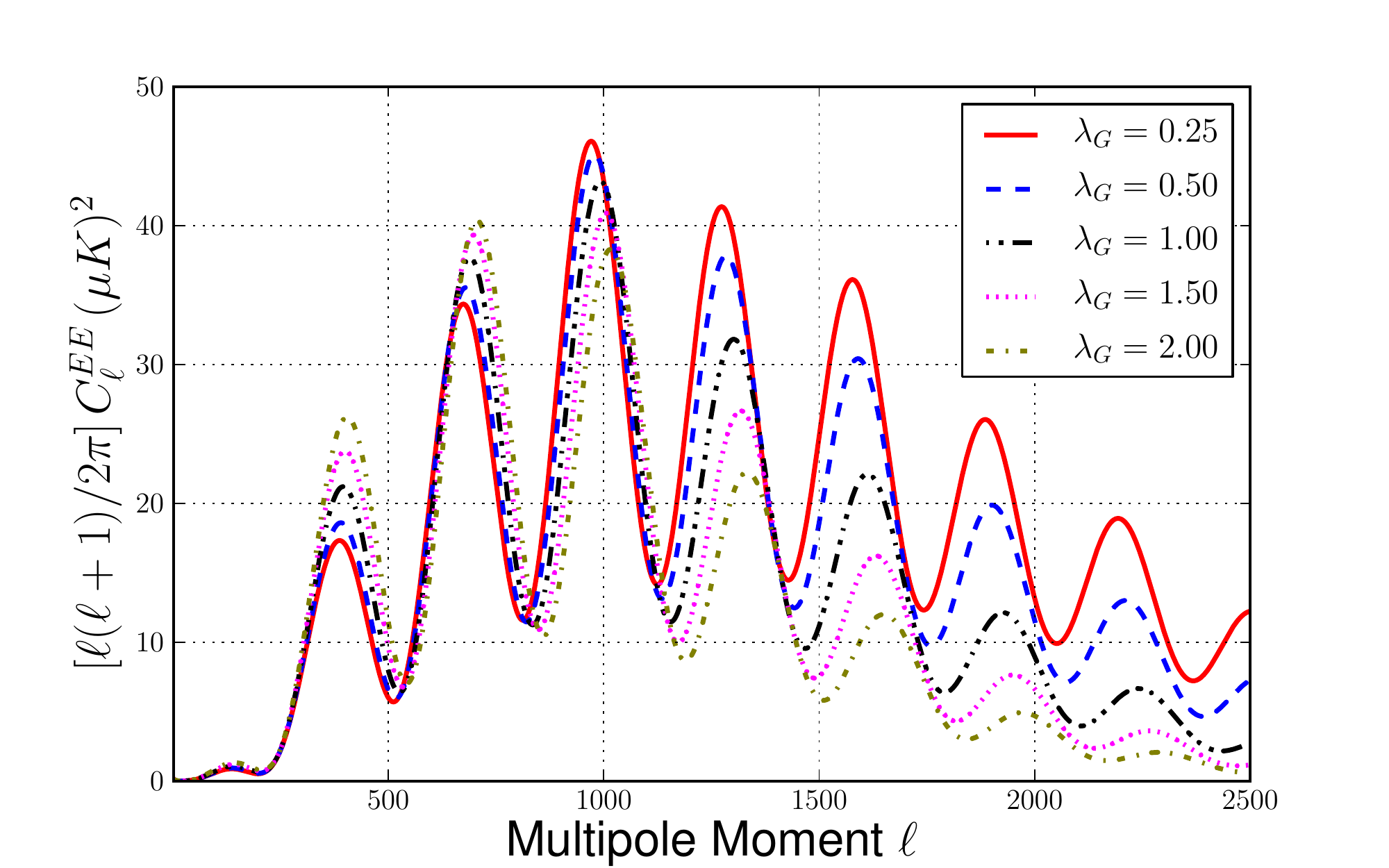}
\includegraphics[width=0.49\textwidth,clip=true]{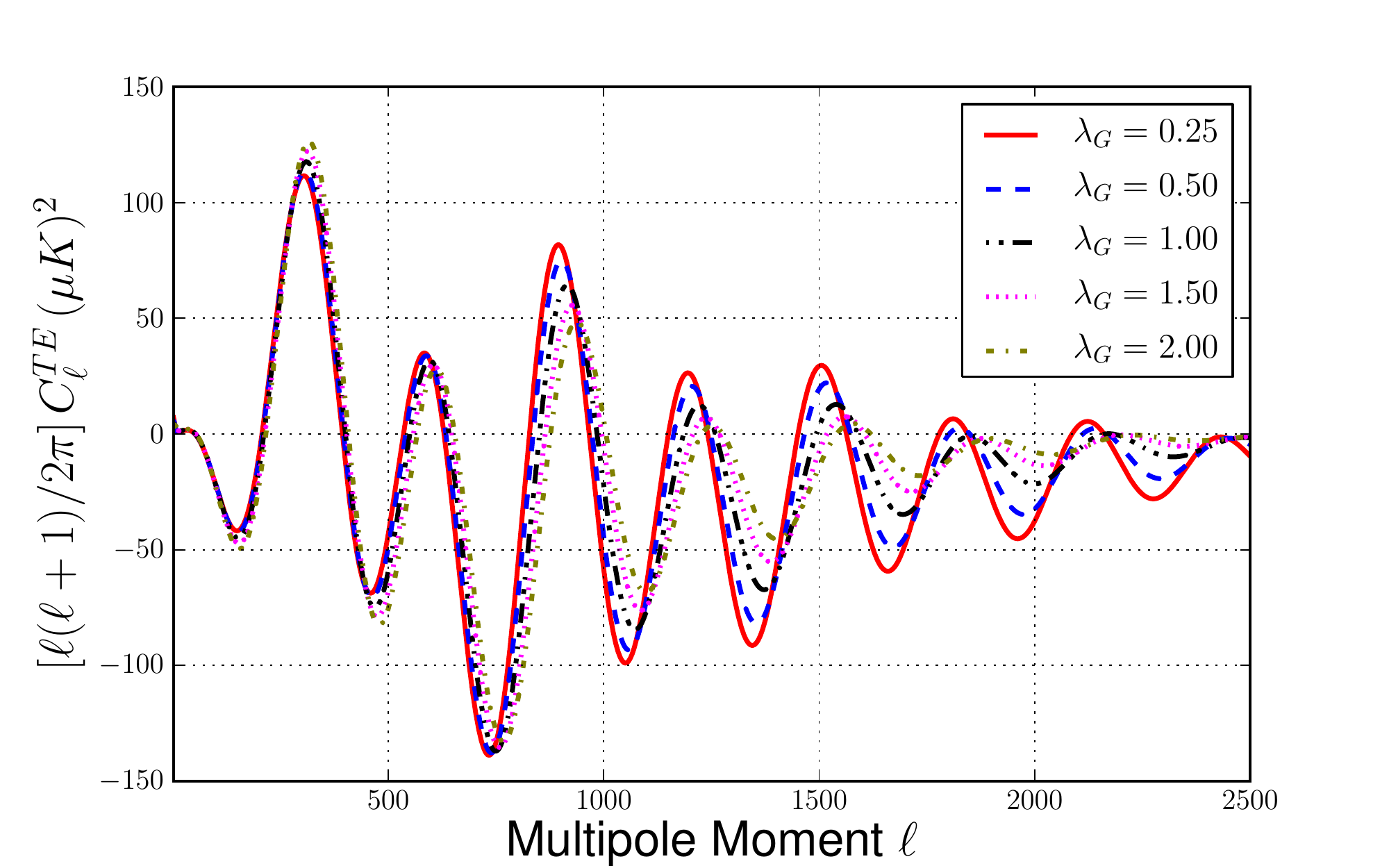}
\caption{The effects of varying $\lambda_{G}$ on the TT, EE, and TE power spectra.}
\label{fig:CMBpower}
\end{center}
\end{figure}

Since the ionization fraction determines the density of free electrons, we also expect the visibility function to change when varying $\lambda_{G}$. The normalized visibility function can be thought as the probability that a photon last scattered at a particular conformal time $\tau$. As $\lambda_{G}$ increases, recombination (and the last scattering of photons) takes place over a longer period of time which means the visibility function becomes broader. This broadening, shown in the right panel of Fig.~\ref{fig:Xe-g}, leads to a damping of anisotropies on small scales on which photons are still scattering. In Fig.~\ref{fig:CMBpower}, we show the net effect of changing $G$ on damping (for $\lambda_{G} > 1$) or enhancement (for $\lambda_{G} < 1$) of the temperature anisotropies, with an emphasis on the small angular scales. Also in Fig.~\ref{fig:CMBpower}, we show the EE and TE power spectra. Since the effects are more dramatic at a small scale or a large $\ell$, we will later use this fact to understand constraints from different experiment data. We also note that the effects from varying $\lambda_G$ have a large correlation with the parameter $n_s$ of the primordial power spectrum, since introducing an appropriate tilt of this spectrum can also act to damp or enhance the small scale peaks.
%

\subsection{Analysis Method}
\vspace*{1mm}
We perform a Markov Chain Monte Carlo analysis using the publicly available Monte Python code~\cite{Audren:2012wb} which interfaces with the CLASS code~\cite{Lesgourgues:2011re,Blas:2011rf}. In addition to $\lambda_{G}$, we sample the concordance $\Lambda$CDM parameters which include the physical baryon and CDM densities, $\omega_{b} = \Omega_{b}h^{2}$ and $\omega_{c} = \Omega_{c}h^{2}$, the Hubble parameter $H_{0}$ at the current time, the scalar spectral index $n_{s}$, the primordial power spectrum normalization $\ln (10^{10} A_{s})$ at $k_0 = 0.05/\mbox{Mpc}$ and the reionization optical depth $\tau$. Flat priors were used on all the above cosmological parameters.

The chains are checked for convergence using the Gelman-Rubin $R-1$ statistic for each parameter~\cite{Gelman:1992zz}. To obtain constraints on the cosmological parameters, the Monte Python package marginalizes over the remaining nuisance parameters. For computing the likelihood we use the package provided by the Planck team with the 2013 data release~\cite{Ade:2013kta}. It contains high and low-$\ell$ TT likelihoods in addition to low-$\ell$ TE and EE likelihoods from WMAP9.  Also included are high-$\ell$ TT likelihoods using 3 years of data from the Atacama Cosmology Telescope (ACT)~\cite{Sievers:2013ica}, the South Pole Telescope (SPT)~\cite{Hou:2012xq} and a Planck lensing likelihood. ACT measures the CMB angular power spectra over a 600 square degree patch of sky at 148 and 218 GHz. SPT does the same measurement over a 800 square degree patch of sky at 95, 150, and 220 GHz. The combined ACT/SPT package covers a multipole moment range $500 < \ell < 3500$ for use in constraining cosmological parameters.

In addition to the Planck package, we use likelihoods for the Baryon Acoustic Oscillations (BAO) and the Hubble Space Telescope (HST) available with the Monte Python code. The HST likelihood comes from~\cite{Riess:2011yx}, which determines the Hubble constant using the Wide Field Camera 3 on the Hubble Space Telescope to observe over 600 Cepheid variables in the host galaxies of 8 recent Type Ia supernovae in the optical and infrared. The dataset covers a redshift range of $0.01 < z < 0.1$. The BAO package contains data from the Sloan Digital Sky Survey (SDSS) (Data Releases 7 and 9)~\cite{SDSS7,SDSS9} and the Six degree Field Galaxy Survey (6dFGS)~\cite{Beutler:2011hx}. These experiments have a mean redshift of $z\simeq 0.10$ ($z \simeq 0.5$) for SDSS Release 7 (9) and $z \simeq 0.05$ for 6dFGS. SDSS Release 7(9) covers 11,663(14,555) square degrees and 6dFGS covers $\sim17,000$ square degrees of sky.

\subsection{Constraints on $G$}
\begin{figure}[th!]
\begin{center}
\hspace*{-0.0cm}
\includegraphics[width=0.6\textwidth]{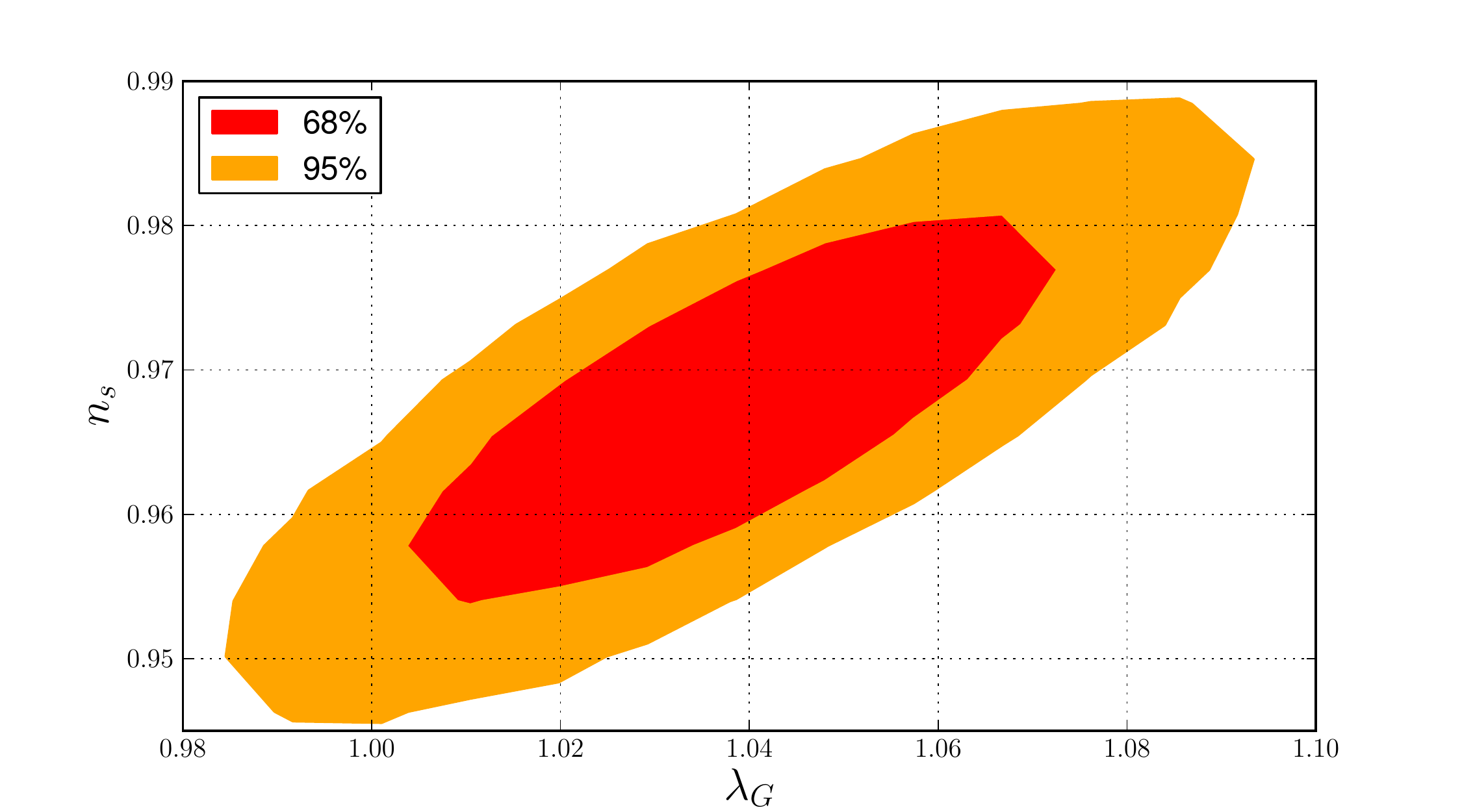}
\caption{68\% and 95\% likelihood contour plots on the $n_{s}$$-$$\lambda_{G}$ plane using the Planck+ACT/SPT+Lensing+BAO+HST dataset.
}
\label{fig:LGvNs}
\end{center}
\end{figure}

\vspace*{1mm}
Using the dataset Planck+ACT/SPT+Lensing+BAO+HST, we report in Table~\ref{table:CoCP} the constraints obtained on the cosmological parameters which were sampled using Monte Python. This dataset provides a constraint on $\lambda_{G}$ at about the $\sim$ 2.2\% level. Including the BBN data raises the mean value of $\lambda_{G}$ and tightens the constraint to the $\sim$ 1.8\% level. To show correlations between $\lambda_{G}$ and some common cosmological parameters, we show the one and two sigma confidence contour plots on the $\lambda_{G}-n_{s}$ planes in Fig.~\ref{fig:LGvNs}. The degeneracy with the scalar spectral index is to be expected since varying $\lambda_{G}$ either delays or hastens recombination, which damps or enhances small angular scale oscillations.

\begin{table}[th!]
\begin{center}
\renewcommand{\arraystretch}{1.3}
\begin{tabular}{|l|c|c|c|c|} 
 \hline \hline
Param & best-fit & mean$\pm\sigma$ & 95\% lower & 95\% upper \\ \hline  \hline
$100~\omega_{b }$ &$2.246$ & $2.236_{-0.03}^{+0.03}$ & $2.176$ & $2.295$ \\ 
$\omega_{\rm cdm }$ &$0.1168$ & $0.1181_{-0.0016}^{+0.0016}$ & $0.1149$ & $0.1214$ \\ 
$H_0$ &$69.74$ & $69.07_{-0.82}^{+0.80}$ & $67.46$ & $70.68$ \\ 
$10^{+9}A_{s }$ &$2.180$ & $2.208_{-0.060}^{+0.052}$ & $2.099$ & $2.321$ \\ 
$n_{s }$ &$0.9685$ & $0.9676_{-0.0091}^{+0.0091}$ & $0.9492$ & $0.9858$ \\ 
$\tau_{\rm reio }$ &$0.087$ & $0.093_{-0.013}^{+0.012}$ & $0.068$ & $0.119$ \\ 
$\lambda_{{G} }$ &$1.033$ & $1.038_{-0.023}^{+0.022}$ & $0.993$ & $1.085$ \\ 
\hline \hline
 \end{tabular} \\ 
 \caption{Constraints on the 7 free cosmological parameters ($\Lambda {\rm CDM} + \lambda_{G}$) using the experimental dataset Planck+ACT/SPT+Lensing+BAO+HST.}
 \label{table:CoCP}
 \end{center}
 \end{table}

We show the constraints on $\lambda_{G}$ from different combinations of datasets in 
Table~\ref{table:datasetVals}. Comparing constraints from different groups, one can see that they are consistent among each other. All of them have the central values to be above unity at a slightly over one sigma confidence level. The datasets of Planck+ACT/SPT with more data at high-$\ell$ provides a strong constraint with the relative error of $\sim 2.4\%$ on $\lambda_G$.  So, we report the cosmological measurement of Newton's gravitational constant using the combination of Planck, ACT/SPT, Lensing, BAO, HST and BBN as
\beqa
G (\mbox{cosmological}) =\lambda_G^2\,G_N (\mbox{CODATA}) = 7.26^{+0.27}_{-0.27}\times 10^{-11}\,\mbox{m}^{3}\mbox{kg}^{-1}\mbox{s}^{-2}  \,,
\eeqa
which has a relative error of $3.7\%$. This cosmologically measured value is roughly consistent with the CODATA and has a 2.2$\sigma$ tension.

\begin{table}[hbt!]
\begin{center}
\renewcommand{\arraystretch}{1.3}
\begin{tabular}{|  c  |  c  | }
  \hline   \hline
  Data    & $\lambda_G$  \\ \hline  \hline     
  Planck & \hspace{2mm} $1.062^{+0.031}_{-0.031}$  \hspace{2mm} \\
  \hline
   \hspace{2mm} Planck+Lensing+BAO  \hspace{2mm} & \hspace{2mm} $1.041^{+0.024}_{-0.027}$  \hspace{2mm} \\
  \hline
  \hspace{2mm} Planck+Lensing+BAO+HST  \hspace{2mm} & \hspace{2mm} $1.046^{+0.026}_{-0.027}$  \hspace{2mm} \\
  \hline
  \hspace{2mm} Planck+Lensing+BAO+BBN  \hspace{2mm} & \hspace{2mm} $1.046^{+0.021}_{-0.021}$  \hspace{2mm} \\
      \hline
     \hspace{2mm} Planck+ACT/SPT  \hspace{2mm} & \hspace{2mm} $1.046^{+0.025}_{-0.028}$  \hspace{2mm} \\
      \hline
     \hspace{2mm} Planck+ACT/SPT+Lensing+BAO+HST  \hspace{2mm} & \hspace{2mm} $1.038^{+0.022}_{-0.023}$   \hspace{2mm} \\
     \hline
          \hspace{2mm} Planck+ACT/SPT+Lensing+BAO+HST+BBN  \hspace{2mm} & \hspace{2mm} $1.043^{+0.019}_{-0.019}$   \hspace{2mm} \\
  \hline    \hline  
\end{tabular}
\caption{Cosmological measurement of $\lambda_{G}$ (with one-sigma errors) from different combinations of data. }
\label{table:datasetVals}
\end{center}
\end{table}

\section{Scalar-mediated Long-range Dark Matter Force}
\label{sec:fifth-force}
The conservative way to differentiate the long-range forces among dark matter and ordinary matter is to introduce an ultra light scalar mediator, which only couples to dark matter particles. There is existing literature on constraining additional Yukawa-like interactions for dark matter~\cite{Friedman:1991dj,Bean:2001ys,Gubser:2004uh,Nusser:2004qu,Bean:2008ac}. Following the notation in Ref.~\cite{Friedman:1991dj}, the interaction Lagrangian for the dark matter and the scalar mediator is
\beqa
{\cal L} \supset \frac{1}{2}\partial_\mu\phi \partial^\mu\phi\,-\,\frac{1}{2}m_\phi^2 \phi^2\,+\, \bar{\chi}i\gamma_\mu \partial^\mu \chi \, - \, \left( 1 + \frac{\phi}{f} \right)\,m_\chi\,  \bar{\chi} \chi \,.  
\eeqa
Here, the ultra-light pseudo Nambu-Goldstone boson $\phi$ could be associated with spontaneous breaking of some global symmetry at the scale $4\pi f$~\cite{Hill:1987bm}. In principle, more complicated terms could enter the $\phi$ potential $V(\phi)$. For our later purpose, the $\phi$ field value is small such that only the leading mass term in the potential is important. For simplicity, we assume that $\phi$ does not couple to the ordinary matter. 

Mediated by the new light scalar field, two dark matter particles with masses $m_{D_1}$ and $m_{D_2}$ have the following static potential 
\beqa
V(r) = - \frac{G_N m_{D_1} m_{D_2}}{r} \left[ 1 + \alpha_f\, e^{-m_\phi \,r}  \right] \,,
\eeqa
where $\alpha_f \equiv M_{\rm pl}^2/f^2$ and $M_{\rm pl} \equiv 1/\sqrt{8\pi G_N} \approx 2.4 \times 10^{18}$~GeV is the reduced Planck scale.

The light scalar field can also contribute to the energy-momentum tensor and therefore can modify the Friedmann equation
\beqa
{\cal H}^{2} = \left( \frac{\dot{a} }{a} \right)^2 =   \frac{8\pi}{3}\,a^2\,G_N\,\left[\rho + \frac{\phi}{f} \,\rho_c + \frac{1}{2}\frac{\dot{\phi}^2}{a^2} + V(\phi) \right]  \,.
\label{eq:fifth-Hubble}
\eeqa
Here, $\rho$ is the total energy density and $\rho_c$ is the cold dark matter energy density. The homogenous scalar $\phi = \phi(\tau)$ has the following equation of motion 
\beqa
\ddot{\phi} + 2\,{\cal H}\,\dot{\phi} + a^{2}\,m_\phi^2\,\phi =  -\frac{1}{f}\,\rho_{c}\,a^{2} \,.
\label{eq:fifth-eom}
\eeqa
%

\subsection{Conditions to Ignore the $\phi$ Background}
\label{sec:ignore-phi-background}
To simplify our discussion for constraining the dark matter fifth force, we will work in the parameter space where we can ignore the $\phi$ contribution to the background evolution ({\it i.e.}, the Friedmann equation). For an ultra-light light scalar satisfying
\beqa
m_{\phi}^{2} \ll \frac{1}{a^{2}}\left( \frac{\ddot{\phi}}{\phi} + 2{\cal H}\frac{\dot{\phi}}{\phi}\right) \,,
\label{eq:scalar-mass-bound}
\eeqa
The solution to the equation of motion in Eq.~(\ref{eq:fifth-eom}) is
\beqa
\dot{\phi}= -3\left(\frac{M_{\rm Pl}}{f}\right)\frac{H_{0}^{2}\Omega_{c}^{0}M_{\rm Pl} }{a^{2}} \ts \int \frac{da}{{\cal H}} \,+\,\frac{c_1}{a^2} \,,
\eeqa
where $c_1$ is an integration constant and $H_0$ and $\Omega_c^0$ are the Hubble parameter and dark matter density at the current time, respectively. Starting from a radiation-dominated universe with ${\cal H}^{2} = H_{0}^{2} \ts \Omega_{R}^{0} \ts a^{-2}$, we have the following solutions (for a particular choice of initial conditions)
\beqa
\phi_{R} &=& -\frac{3}{2}\left(\frac{M_{\rm Pl}}{f}\right)\frac{\Omega_{c}^{0}}{\Omega_{R}^{0}} M_{\rm Pl}\ts a
\,, \\
\dot{\phi}_{R} &=& -\frac{3}{2}\left(\frac{M_{\rm Pl}}{f}\right)\frac{\Omega_{c}^{0}}{\sqrt{\Omega_{R}^{0}}} H_{0} M_{\rm Pl} \,.
\label{eq:fifth-background-rad}
\eeqa

In the matter dominated region with ${\cal H}^{2} = H_{0}^{2} \ts \Omega_{M}^{0} \ts a^{-1}$, we find the following solutions
\beqa
\phi_{M} &=& -2\left(\frac{M_{\rm Pl}}{f}\right)\frac{\Omega_{c}^{0}}{\Omega_{M}^{0}}\,M_{\rm Pl} \, \left[ \ln(a)- \frac{2}{3}\frac{C_{1}^{M}}{a^{3/2}}+ C_{2}^{M}\right]
\,, \\
\dot{\phi}_{M} &=& -2\left(\frac{M_{\rm Pl}}{f}\right)\frac{\Omega_{c}^{0}}{\sqrt{\Omega_{M}^{0}}} \,H_{0} M_{\rm Pl} \,\left( a^{-1/2}+ \frac{C_{1}^{M}}{a^{2}}\right)  
\,.
\eeqa
The two integration constants $C_1^M$ and $C_2^M$ can be determined by requiring both $\phi$ and $\dot{\phi}$ to be continuous functions: $\phi_R(a_{\rm eq}) = \phi_M(a_{\rm eq})$ and $\dot\phi_R(a_{\rm eq}) = \dot\phi_M(a_{\rm eq})$. The equality of matter and radiation happens at $a_{\rm eq} \approx 1/3600$, which determines the integration constants
\beqa
C_1^M\,=\,-\,\frac{1}{4} \, a_{\rm eq}^{3/2} \approx   - 1.16\times 10^{-6} \,, \qquad 
C_2^M\, =\, \frac{7}{12} - \ln{a_{\rm eq}} \approx 8.77 \,.
\eeqa

To ignore the $\phi$ background contributions to the Hubble parameter in Eq.~(\ref{eq:fifth-Hubble}), we need to have all three $\phi$-related terms to be suppressed. It turns out that the requirement, $\phi\,\rho_c /f < \rho$, in the matter-dominated era provides the most stringent bound and requires
\beqa
f \gg \sqrt{2\,C_2^M}\,M_{\rm Pl} \approx 4.2 \,M_{\rm Pl} \,,\qquad \mbox{or} \qquad \alpha_f \ll 0.06 \,.
\label{eq:fifth-alpha-constraint}
\eeqa
The bound on the scalar mass from Eq.~(\ref{eq:scalar-mass-bound}) can be simplified to be
\beqa
\frac{m_\phi}{H_0} \ll \sqrt{\frac{\Omega^0_M}{C_2^M}}  \approx 0.2 \,,
\label{fifth-mass-constraint}
\eeqa
or $m_\phi  \ll {\cal O}(10^{-34}~\mbox{eV})$. In the following numerical analysis, we will stay in the parameter space satisfying Eq.~(\ref{eq:fifth-alpha-constraint}) and Eq.~(\ref{fifth-mass-constraint}). 

\subsection{Linear Perturbation Equations}
\label{sec:fifth-force-linear}
To derive the linear perturbation equations, we expand the scalar field into
\beqa
\phi(x, \tau) \,=\, \phi_0(\tau) \,+\, \varphi(x, \tau) \,,
\eeqa
with $\phi_0(\tau)$ as the background field and $\varphi(x, \tau)$ as the first order perturbation function. Perturbing the $\phi$ equation of motion, we arrive at the following equation for the $\varphi$ evolution~\cite{Bean:2001ys}
\beqa
\ddot{\varphi} + 2\,{\cal H}\,\dot{\varphi} + (k^{2} + a^{2}m_{\phi}^{2})\varphi = -\frac{1}{2}\dot{h}\,\dot{\phi_0} - \frac{1}{f}\,\rho_{c}\,\delta_{c}\,a^{2} \,.
\label{eq:fifth-varphi-eom}
\eeqa
Here, the function $h(k, \tau)$ is the scalar metric perturbation in the synchronous gauge~\cite{Ma:1995ey}. The new second-order differential equation in Eq.~(\ref{eq:fifth-varphi-eom}) requires two initial conditions for the $\varphi$ field. Deep in the radiation epoch, keeping the leading terms in powers of $a$ and neglecting $k^2$ and $m_\phi^2$ terms, we have $ (d/d\tau)(a^{2}\dot{\varphi}) = - \frac{1}{2}\, \dot{h}\,\dot{\phi}_0\,a^2$. Using the scaling solutions of $h \propto \tau^2$ and $a \propto \tau$ and noticing that $\dot{\phi}_0$ is a constant from Eq.~(\ref{eq:fifth-background-rad}), we have the following initial conditions 
\beqa
\varphi = -\frac{1}{12}h\,\dot{\phi}_0\,\tau  \,, \qquad \qquad 
\dot{\varphi} = -\frac{1}{4}h\,\dot{\phi}_0 \,.
\eeqa

The first-order perturbed Einstein equations are 
\beqa
k^{2}\,\eta - \frac{1}{2}{\cal H}\,\dot{h} &=& 4\pi G_N\,a^{2}\,\delta T^{0}_{\;\;0}  \,, \\
k^{2}\,\dot{\eta} &=& 4\pi G_N\,a^{2}\,ik^{i} \delta T_{\;\;i}^{0}    \,, \\
\ddot{h} +2\,{\cal H}\,\dot{h} - 2k^{2} \eta &=& -8\pi G_N \, a^{2} \, \delta T_{\;\;i}^{i}   \,, \\
\ddot{h} + 6\ddot{\eta} + 2\,{\cal H}\,(\dot{h} + 6\dot{\eta}) - 2k^{2}\,\eta &=& 24\pi G_N\, a^{2} (\hat{k}_{i}\hat{k}_{j} - \frac{1}{3}\delta_{ij})(T^i_{\ts \ts j} - \delta^{i}_{\ts \ts j} T^k_{\ts \ts k}/3)  \,.
\eeqa
The new contributions to the energy momentum tensor from the scalar field are calculated to be 
\beqa
\delta T_{\;\;0}^{0}(\phi) &=&  -\frac{\rho_{c}}{f}\left[ \varphi \,+\, \phi_0\, \delta_{c} \right] \,-\, \frac{\dot{\varphi}\,\dot{\phi}_0 }{a^{2}}\,-\, m_{\phi}^{2}\,\phi_0\,\varphi    \,, \\
ik^i \delta T_{\;\;i}^{0}(\phi) &=& -\frac{1}{a^{2}}\dot{\phi}_0\,k^{2}\varphi  \,, \\
\delta T_{\;\;i}^{i}(\phi) &=&  \frac{3}{a^{2}}\dot{\phi}_0\,\dot{\varphi} \,-\, 3\,m_{\phi}^{2}\,\phi_0\,\varphi    \,,
\eeqa
which agree with the formulas in Ref.~\cite{Bean:2001ys}. The dark matter density perturbation still obeys $\dot\delta_{c} = -\frac{1}{2}\dot{h}$, simply from the conservation of energy-momentum. Since the field $\phi_0$ and $\varphi$ change the values of $h$ through the Einstein equations, the dark matter density perturbation is affected by $\phi_0$ in this indirect way.

\subsection{Constraints on $\alpha_f \equiv M_{\rm pl}^2/f^2$}
\label{sec:fifth-constraints}
\begin{figure}[th!]
\begin{center}
\hspace*{-0mm}
\includegraphics[width=0.47\textwidth,clip=true,viewport= 20 0 590 340]{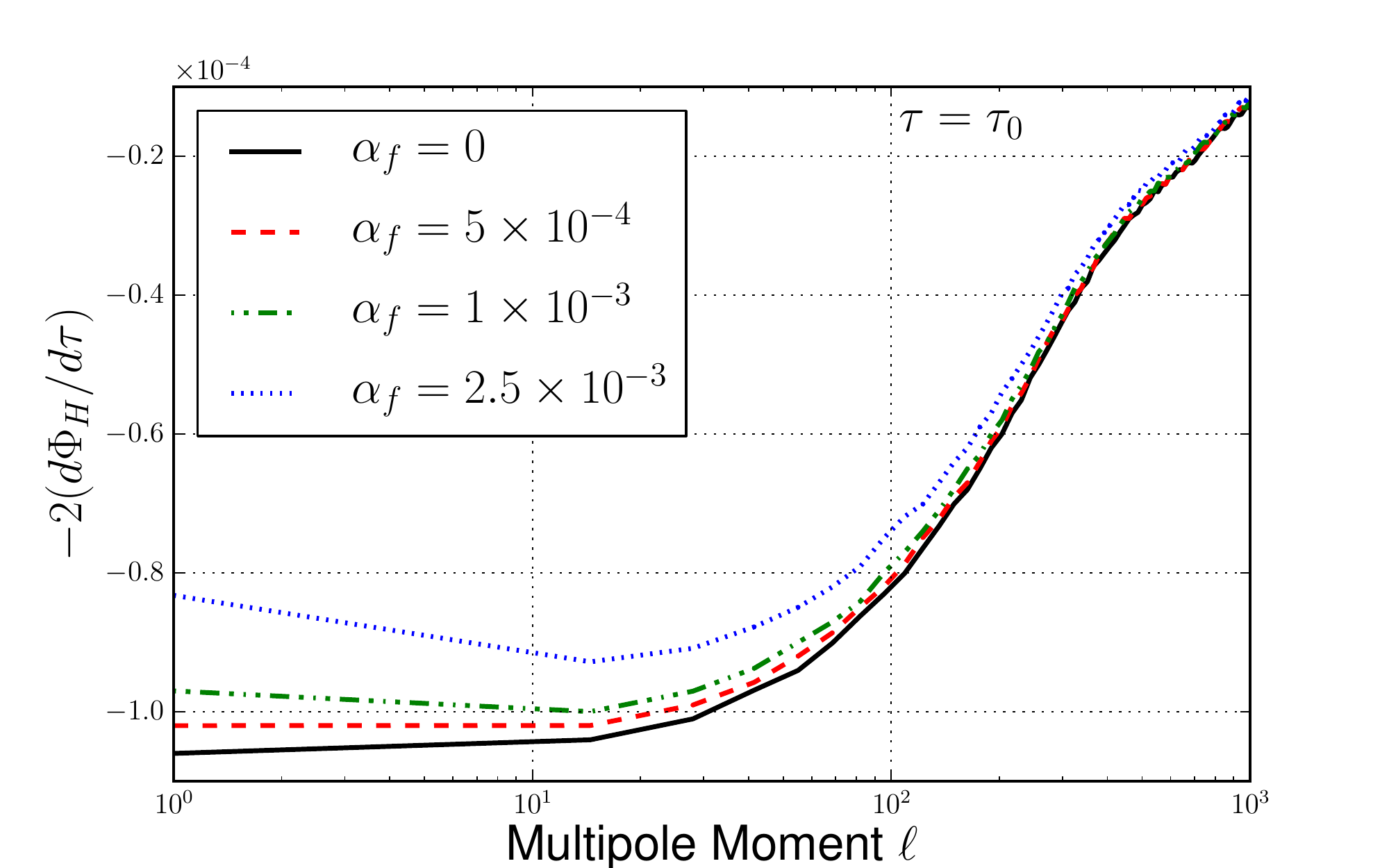} \hspace{3mm}
\includegraphics[width=0.48\textwidth,clip=true,viewport= 10 0 590 340]{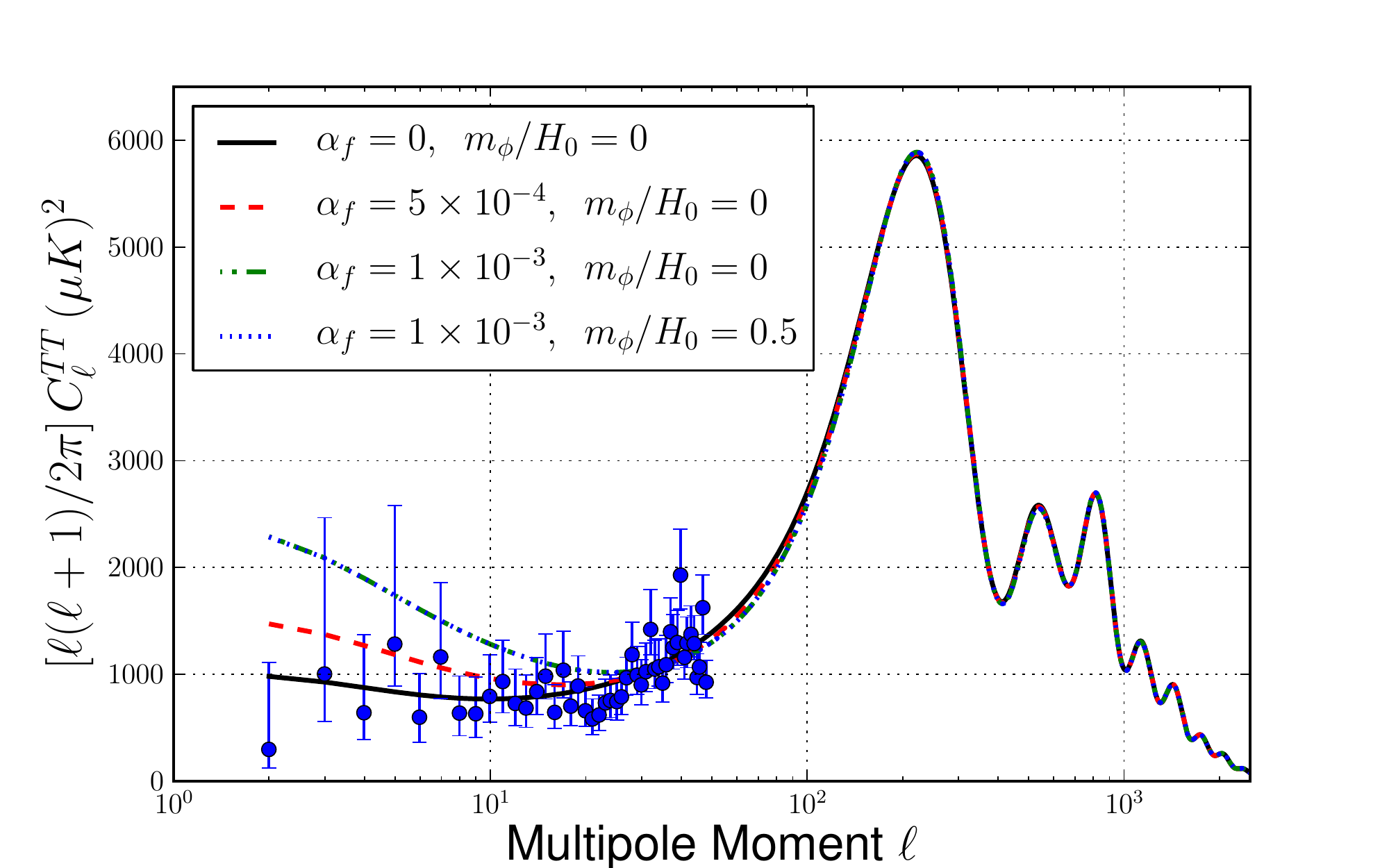}
\caption{{\bf Left panel:} the derivative of the gauge-invariant scalar metric perturbation as a function of multipole moments, $\ell$, for different values of the fifth force strength $\alpha_f$ and at the current time. {\bf Right panel:} the CMB TT power spectra in $\ell$ for different values of $\alpha_f$ and $m_\phi$. The curves for $m_\phi/H_0=0$ and $m_\phi/H_0=0.5$ are overlapping each other. Also shown here is the Planck three year data points for $\ell \leq 50$~\cite{Ade:2013kta}.}
\label{fig:fifth-TT}
\end{center}
\end{figure}

 Because of our choice of initial conditions for the scalar mediator evolution, the main constraints on the strength of the fifth force come from the perturbation part rather than the background evolution. The CMB temperature anisotropy turns out to provide the most stringent constraint. As already mentioned in Ref.~\cite{Koivisto:2005nr,Morris:2013hua}, the additional long-range attractive force among dark matter particles can introduce the late time variation of the gravitational potential. The ISW effect acts to increase the power for low multipoles. To confirm this observation, we show the conformal time-derivative of the gravitational potential in the left panel of Fig.~\ref{fig:fifth-TT}. Since the universe is expanding, the magnitude of the potential is decaying in time and as such the derivative is negative. Comparing to the ordinary cosmology with $\alpha_f=0$ and from Fig.~\ref{fig:fifth-TT}, one can see that increasing the value of $\alpha_f$ leads to a larger gravitational potential and a more dramatic ISW effect. 

In the right panel of Fig.~\ref{fig:fifth-TT}, we show the CMB temperature anisotropies in terms of the multipole moments. Since the ISW effect mainly affects the small $\ell$ region, increasing the value of $\alpha_f$ leads to a larger power spectrum. Within the validity of our approximation of ignoring the scalar field contribution to the background evolution, the difference between a massless $m_\phi/H_0=0$ and a small mass such as $m_\phi/H_0=0.5$ is negligible. So, we do not anticipate that the CMB temperature anisotropies can constrain $m_\phi$ in the regime where the approximation condition in Eq.~(\ref{fifth-mass-constraint}) is satisfied.

\begin{figure}[th!]
\begin{center}
\vspace*{0.5cm}
\includegraphics[width=0.51\textwidth,viewport= 20 0 600 370]{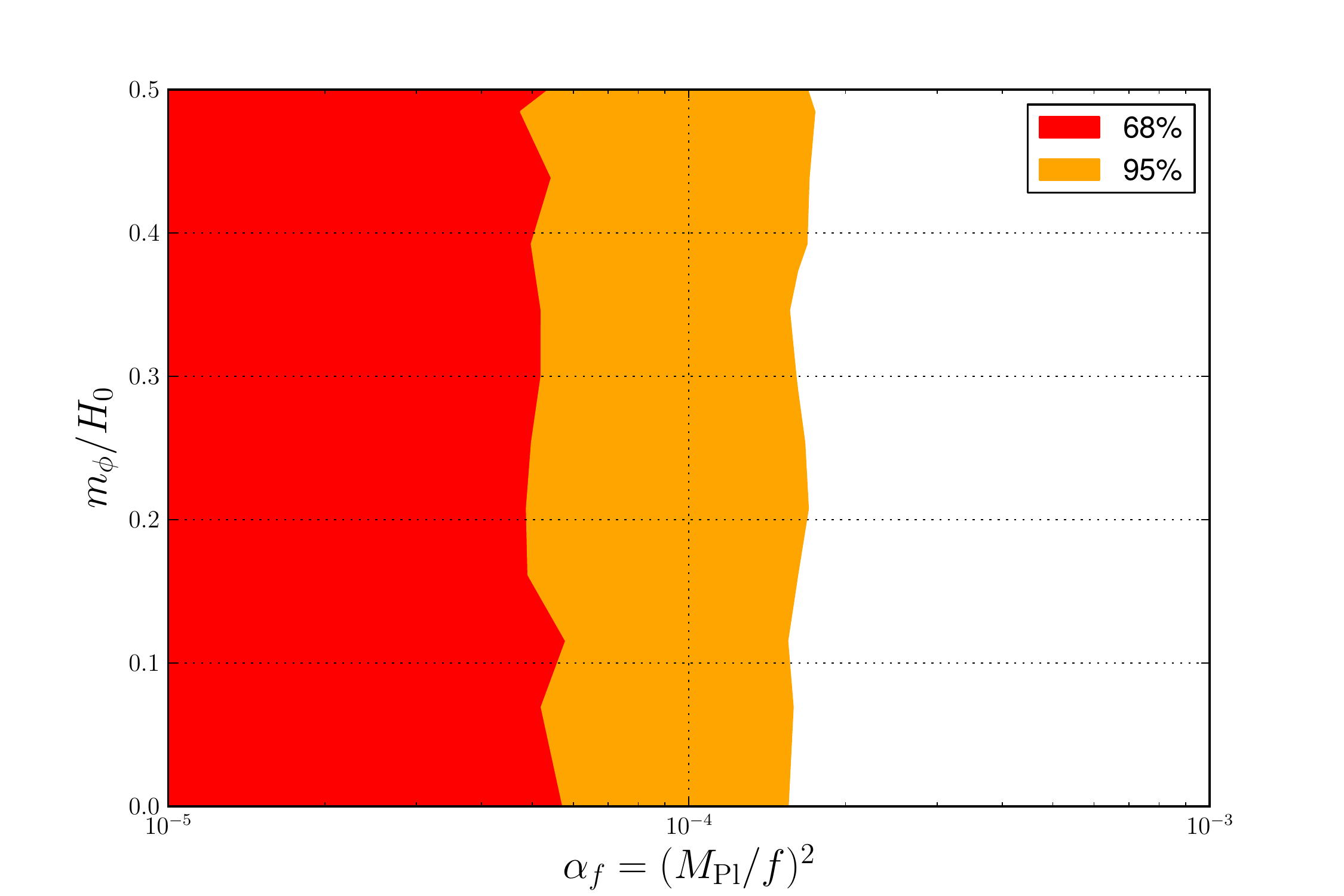}
\caption{The constraints on the fifth force strength $\alpha_f$ and the mediator mass $m_\phi$ from the Planck+Lensing data.
 }
\label{fig:fifth-contour}
\end{center}
\end{figure}
In Fig.~\ref{fig:fifth-contour} and using the data from Planck+Lensing, we show the cosmological constraints on the fifth force strength and the mediator mass. One can see that the data is insensitive to the mediator mass, but does provide a stringent bound on the fifth force strength as 
\beqa
\alpha_f \equiv M_{\rm pl}^2 / f^2  \le 1.62\times 10^{-4}\,\quad (95\%~\mbox{C.L.}) \,.
\eeqa 
Compared to the results in Ref.~\cite{Bean:2008ac}, which used the WMAP 5-year data~\cite{Nolta:2008ih}, our results using the Planck data show a dramatic improvement on constraining an additional long-range force among dark matter.

\section{Dark Matter Weak Equivalence Principle: Two-Fluid Description}
\label{sec:two-fluid}
In this section, we use cosmological data to constrain the difference between inertial and gravitational mass for the dark matter. This immediately requires a violation of WEP. For baryonic matter, modern torsion-balance experiments report that the difference between inertial and gravitational masses is zero at the $10^{-13}$ level~\cite{Wagner:2012} (see also~\cite{Adelberger:2003zx,Adelberger:2006dh}). Thus, violations of the WEP in the visible sector are tightly constrained. Therefore, in this section we consider a violation of the WEP in the dark matter sector only. We introduce a new parameter $\lambda_D$ such that the ratio of dark matter gravitational to inertial mass is given by
$\lambda_{D} =  m_{D}^{\rm grav}/m_{D}$ with $m_D$ as the inertial mass. For baryonic particles, the gravitational mass is assumed to be the same as the inertial mass. For two baryonic particles $b_1$ and $b_2$ and two dark matter particles $D_1$ and $D_2$,  the gravitational forces in terms of the particle inertial masses are 
\beqa
F_{b_1, b_2} = - \frac{G_N m_{b_1} m_{b_2}}{r^2}, \quad   F_{b_i, D_j} = - \lambda_D\frac{G_N m_{b_i} m_{D_j}}{r^2}, \quad F_{D_1, D_2} = - \lambda_D^2\frac{G_N m_{D_1} m_{D_2}}{r^2} \,. 
\eeqa
There is a simple relation, $F_{b_1, b_2} F_{D_1, D_2} = F_{b_1, D_1} F_{b_2, D_2}$, which does not hold in section~\ref{sec:fifth-force} where there is no ``fifth force'' between one dark matter and one baryonic particles.

As pointed out in the textbook~\cite{Mukhanov-book},  one can derive parts of the Friedmann and linear perturbation equations of cosmology using a post-Newtonian description. The Friedmann equation can be derived by considering a sphere filled with a homogenous and isotropic matter density distribution and studying the radius change as a function of time. For pressureless matter, one can use the continuity and Euler equations of fluid mechanics to match to the energy-momentum conservation equations from general relativity. In our case with different Newtonian forces for dark and ordinary matter, we use this post-Newtonian language to motivate modifications to the cosmological background and linear perturbation equations. Following the same approach as  Ref.~\cite{Mukhanov-book} takes with a single scale factor for all matter, we derive two coupled ``Friedmann equations'' which govern the expansion of dark and ordinary matter as two coupled fluids, each with their own scale factor. We would like to note that we are using this post-Newtonian, non-relativistic description in order to motivate a phenomenological parameterization of dark matter WEP breaking, since true WEP breaking is unlikely to be realized within the framework of general relativity. A full, generally covariant model that may encode these deviations is beyond the scope of this work.

The classical picture for our two-fluid description is two expanding spheres with the centers located at the same point. The mass densities, $\rho_b$ and $\rho_D$, for both fluids are assumed to be uniform inside the spheres. For a probing baryonic matter particle with mass $m_b$ on the surface of a sphere with radius $r_b$, its motion is governed by the amount of baryon and dark matter inside the $r_b$ radius sphere
\beqa
m_{b}\,\frac{d^2{r}_{b}}{dt^2}=-\frac{G_N\,m_{b}\,M_b(r_b)}{r_{b}^{2}}\,-\lambda_{D} \,\frac{G_N\,m_{b}\,M_{D}(r_b)}{r_{b}^{2}}\,,
\label{eq:newt_baryon_eom}
\eeqa
where $M_b(r) \equiv 4\pi r^3 \rho_b /3$ and $M_D(r) \equiv 4\pi r^3 \rho_D /3$. Here, to describe the particle motion, all masses $m_b$, $M_b$ and $M_D$ are inertial masses. Similarly, for a probing dark matter particle with mass $m_D$ and at a radius $r_D$, one has the equation of motion to be
\beqa
m_{D}\,\frac{d^2{r}_{D}}{dt^2}=-\lambda_D\,\frac{G_N\,m_{D}\,M_b(r_D)}{r_{D}^{2}}\,-\lambda_{D}^2 \,\frac{G_N\,m_{D}\,M_{D}(r_D)}{r_{D}^{2}}\,. 
\label{eq:newt_dark_eom}
\eeqa
In general, the two radius functions in time $r_b(t)$ and $r_D(t)$ could be independent of each other. So, to match to the Friedmann equations, we need to introduce two scale factors, $a(t)$ and $a_D(t)$, for baryonic matter and dark matter, respectively.  

Requiring the radii proportional to the scale factors, we have $r_b(t) = r^0_b a(t)/a^0$ and $r_D(t) = r^0_D a_D(t)/a^0_D$. We then rewrite Eqs. (\ref{eq:newt_baryon_eom})(\ref{eq:newt_dark_eom}) as 
\beq
\frac{1}{a}\frac{d^2a}{dt^2}=-\frac{4\pi G_N}{3}\big[\rho_{b}(a)\, +\, \lambda_{D} \ts \rho_{D}(a_D)\big] \,, \quad 
\frac{1}{a_D}\frac{d^2a_D}{dt^2}=-\frac{4\pi G_N}{3}\big[ \lambda_{D} \ts \rho_{b}(a)\, +\,\lambda_{D}^{2} \ts \rho_{D}(a_D) \big] \,.
\label{eq:DFE}
\eeq
Here, the densities are diluted with the expansion of the sphere and are $\rho_b = \rho^0_b [a^0/a(t)]^3$ and $\rho_D = \rho^0_D [a^0_D/a_D(t)]^3$.  The equations in Eq.~(\ref{eq:DFE})  describe the acceleration of baryonic and dark matter particles, when the dark matter sector violates the WEP. The power of $\lambda_{D}$ corresponds to the number of dark matter masses present in the interaction, {\it e.g.} a dark matter particle moving in response to a baryonic source receives one power of $\lambda_{D}$. We see that in the limit where the WEP is restored ($\lambda_{D} \rightarrow 1$), both equations reduce to the usual Friedmann equation as expected.  Extending the above analysis to include the radiation energy density $\Omega_R$ and dark energy density $\Omega_\Lambda$ by assuming they couple to gravity in the same way as the baryons, we have our final Friedmann equations
\beqa
\frac{1}{a}\frac{d^2a}{dt^2} &=& -\frac{\widetilde{H}_{0}^{2}}{2}\left[ 
\frac{\Omega_{b}}{a^{3}} + \frac{2\Omega_{R}}{a^4} +(1+3w)\frac{\Omega_{\Lambda}}{a^{3w+3}}
+ \frac{\lambda_{D} \Omega_{D}}{a_{D}^{3}}\right] \,,
\label{eq:TBFE} \\
\frac{1}{a_D}\frac{d^2a_D}{dt^2} &=& -\frac{\widetilde{H}_{0}^{2}}{2}\left[ \lambda_{D}\left(
\frac{\Omega_{b}}{a^{3}} + \frac{2\Omega_{R}}{a^4} +(1+3w)\frac{\Omega_{\Lambda}}{a^{3w+3}} \right)
+ \frac{\lambda^{2}_{D} \Omega_{D}}{a_{D}^{3}}\right]\,.
\label{eq:TDFE}
\eeqa
Here, we introduce the parameter $\widetilde{H}_0$ to draw a distinction from the ordinary Hubble constant, $H_0 = \frac{1}{a}\frac{da}{dt}|_{\rm today}$, at the current universe. For $\lambda_D=1$ with a single scale factor, we have $\widetilde{H}_0 = H_0$. For the ordinary $\Lambda$CDM cosmology with a flat universe, the dark energy density is not an independent parameter and is given by $\Omega_\Lambda = 1 - \Omega_b - \Omega_R - \Omega_D$. For our case with two scale factors, we will keep $\Omega_\Lambda$ as a free parameter.  

Compared to ordinary cosmology with a single scale factor, we have one additional second-order differential equation, which requires two more initial conditions. A simple way to fix the dark matter scale factor initial conditions is to have $a_D = a$ and $da/dt = da_D /dt$ at some time $a=a_{\rm init}$. We assume that the dark WEP breaking is turned on at this time, parametrized by $z_{T} = a^{-1}_{\rm init} - 1$. Physically, this transition redshift $z_{T}$ corresponds to a scale where some interaction coupling dark matter to the Standard Model falls out of equilibrium. Before the transition redshift $z_{T}$, everything evolves as a one-component fluid described by a single scale factor $a$. After $z_{T}$, the dark matter decouples from the other components and evolves as a separate fluid according to a dark scale factor $a_{D}$. 

\begin{figure}[th!]
\begin{center}
\hspace*{-0.0cm}
\includegraphics[width=0.6\textwidth]{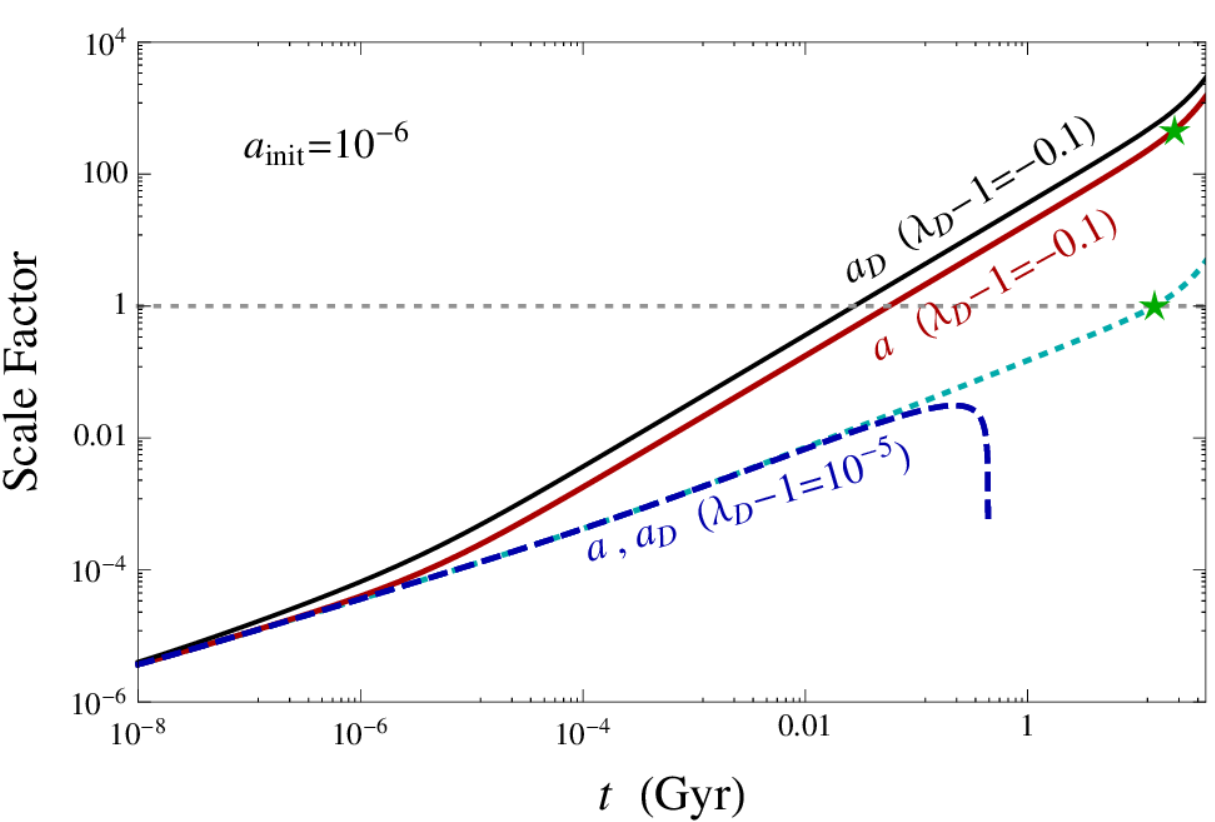}
\caption{The ordinary matter and dark matter scale factors as a function of time. The cyan and dotted line is for ordinary cosmology with $\lambda_D = 1$. The two green five-stars correspond to the time with the current observed Hubble constant.}
\label{fig:scale-factor}
\end{center}
\end{figure}

Using the measured cosmological parameters from Planck~\cite{Ade:2013zuv}, we show the behaviors of the two scale factors as a function of time in Fig.~\ref{fig:scale-factor}. 
In this plot, we choose $a_{\rm init} = 10^{-6}$ and use the equations of ordinary cosmology to calculate $da/dt$ at the time corresponding to $a_{\rm init} = 10^{-6}$. We then treat this as the initial condition to solve the coupled equations in Eqs.~(\ref{eq:TBFE})(\ref{eq:TDFE}) for two different $\lambda_D=1-0.1$ and $1+10^{-5}$. As can be seen from Fig.~\ref{fig:scale-factor} and for the $\lambda_D < 1$ case (red and black lines), both scale factors increase faster than the standard cosmology with $\lambda_D=1$. The dark matter scale factor increases faster than the ordinary matter one. This is due to the fact that less matter leads to an open universe and the effective matter for the $a_D$ equation of motion is smaller than the one for $a$. For the $\lambda_D > 1$ case, a closed universe will be obtained even before one obtains the current measured Hubble constant. Therefore, we generically anticipate more stringent constraints for the $\lambda_D > 1$ case than the $\lambda_D < 1$ case.  For a larger value of $a_{\rm init}$ or a smaller value of $z_T$, the difference between the new scale factor and the standard one become smaller. This is simply because the dark matter fluid has less time to evolve decoupled from the baryonic matter fluid. 

\subsection{Dependence of CMB Anisotropy on $\lambda_D$}
So far we have only motivated modifications to the background evolution equations. To motivate modifications to the linear perturbation equations for the dark matter fluid, we will work in the $\it baryon$ co-moving frame and define the conformal time using the ordinary baryon scale factor $dt = a(\tau)d \tau$. The linear perturbation equations for ordinary baryons stay the same. The linear perturbations for the dark matter fluid receive the following modifications (see Appendix~\ref{app:linear-equations} for a detailed explanation)
\beqa
&&\dot{\delta}_{D}\,+\,({\cal H}- {\cal H}_D)\,( 3 + k \ts \partial_{k})\,\delta_{D} \,=\,3\,\delta\dot{\phi}-\theta_{D} \,, \label{eq:linear-delta-D} \\
&&\dot{\theta}_{D}\,+\,{\cal H}\,\theta_D \,+\,2\,({\cal H}- {\cal H}_D)\,( 1 + k \ts \partial_{k} )\,\theta_{D} \,=\,  k^{2}\,\delta\psi\,, \label{eq:linear-theta-D} 
\eeqa
in the conformal Newtonian gauge. We follow the notation used in Ref.~\cite{Ma:1995ey} with dots indicating conformal time derivatives and $\delta_D \equiv (\rho_D - \rho_D^0)/\rho_D^0$ and $\theta_D \equiv \nabla \cdot \delta {\bf v}_D$. Here, ${\cal H}\equiv \frac{1}{a}\frac{da}{d\tau}$ and ${\cal H}_D\equiv \frac{1}{a}\frac{da_D}{d\tau}$. Because the dark matter co-moving frame is not identical to the baryon one, additional bias terms proportional to $({\cal H} - {\cal H}_D)$ enter the above equations. The perturbed Poisson equation, in the {\it baryon} co-moving frame, is also modified as
\beqa
&&k^2\,\delta\phi + 3 {\cal H}\,( \delta \dot{\phi} + {\cal H} \,\delta\psi)= - 4\pi G_N\,a^{2} \left[\delta{\rho}\,+\,(\lambda_D-1)\,\rho_{D}^{0}\,\delta_{D} \right]  \,.
\eeqa

\begin{figure}[th!]
\begin{center}
\hspace*{-0.0cm}
\includegraphics[width=0.49\textwidth]{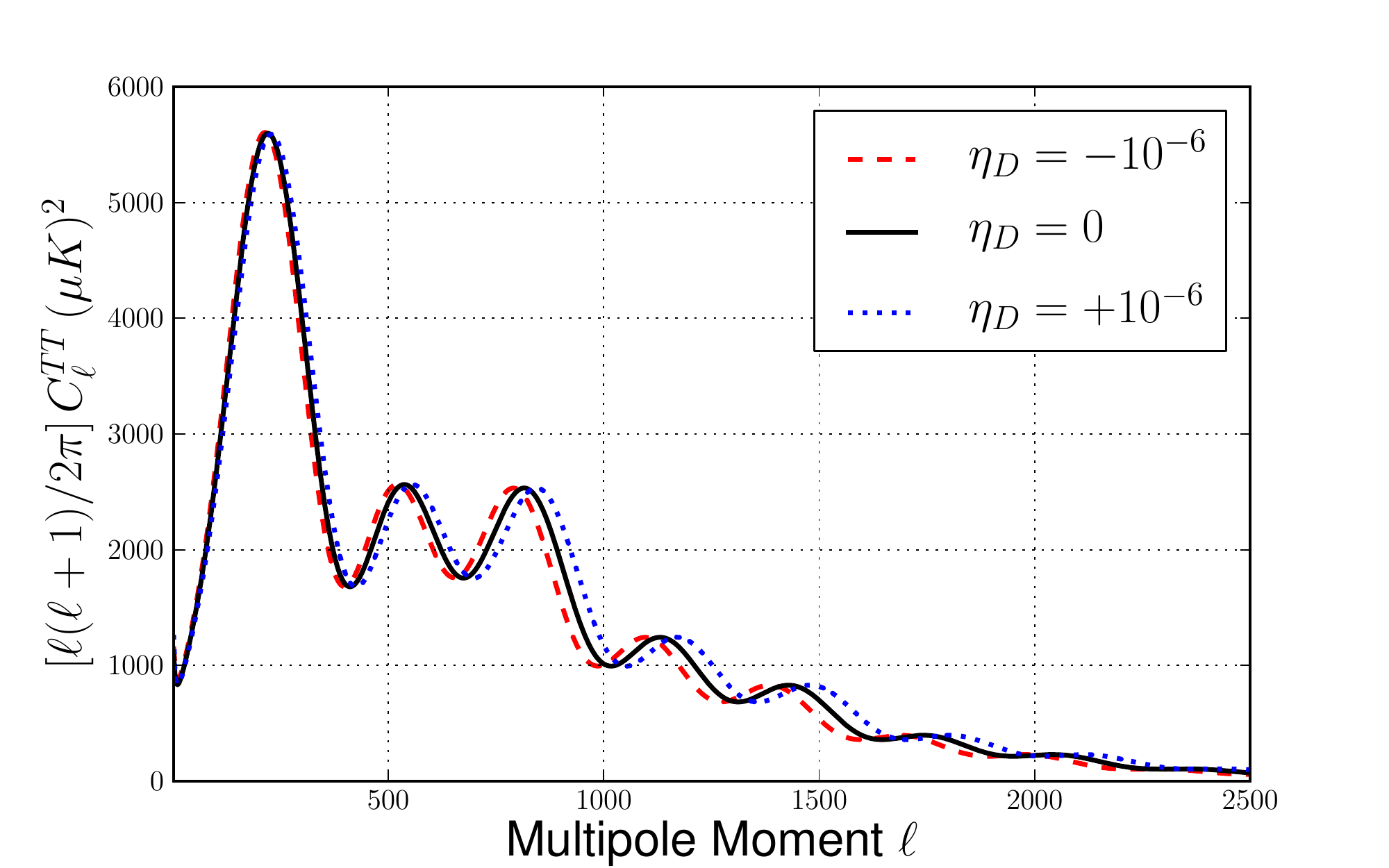}
\includegraphics[width=0.49\textwidth]{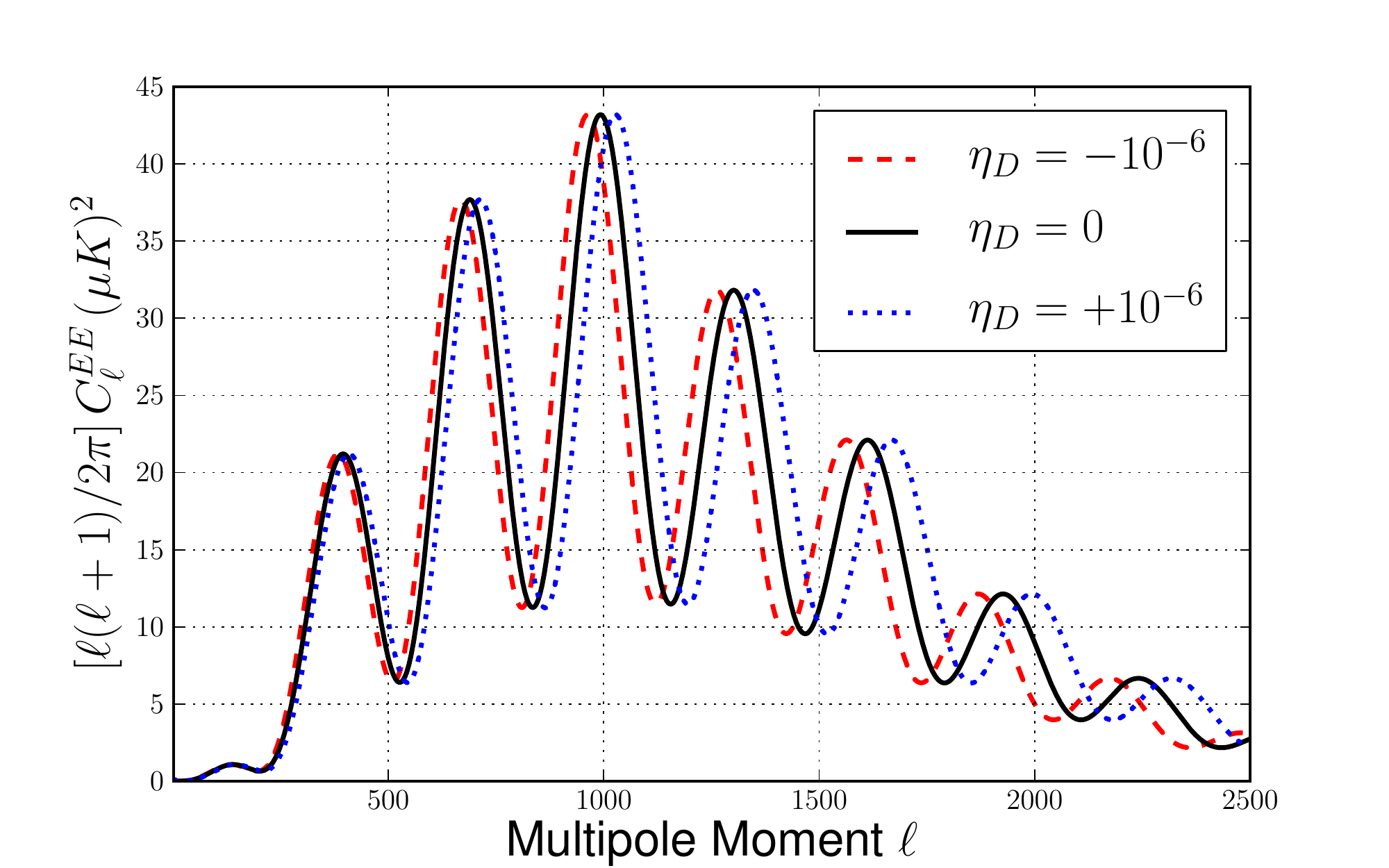}
\includegraphics[width=0.49\textwidth]{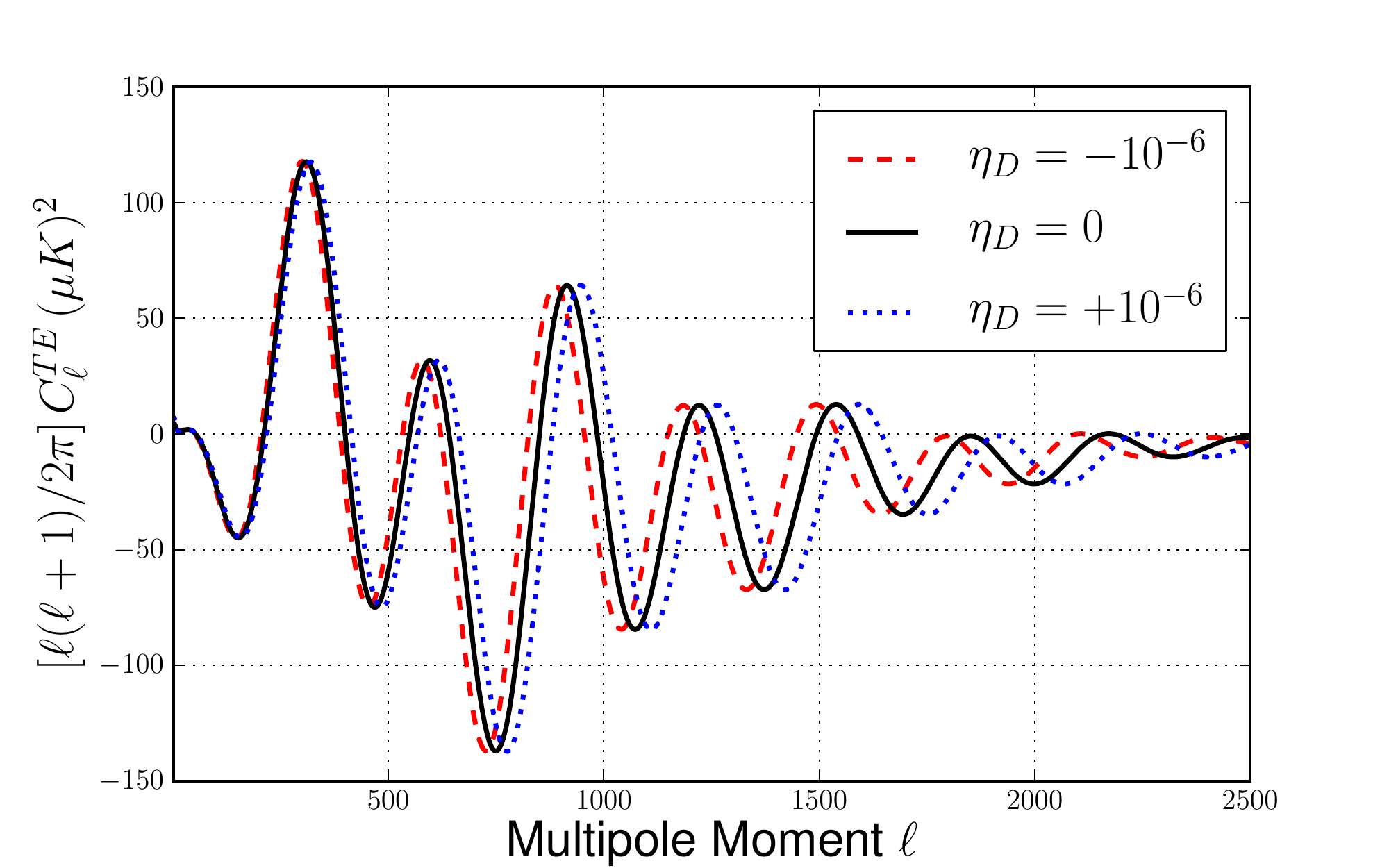}
\caption{The effect of varying $\eta_{D}\equiv\lambda_{D} - 1$ on the TT, EE, and TE power spectra for a transition redshift of $z_{T} = 10^{5}-1$.}
\label{fig:CMB_DM_power}
\end{center}
\end{figure}

We implement these modifications in the CLASS code~\cite{Lesgourgues:2011re,Blas:2011rf} and show the effects of dark matter WEP breaking on the CMB power spectra in Fig.~\ref{fig:CMB_DM_power}. 
For these plots, we fix the transition redshift to be $z_T = 10^5 -1$. For even a small deviation of $\lambda_D$ from one (or a tiny $\eta_D\equiv \lambda_D - 1$), dark matter WEP breaking has visible effects on the CMB power spectra. From Fig.~\ref{fig:CMB_DM_power}, one can see that the main effect is to shift the location of peaks and troughs without changing the height of the peaks. This effect has some similarity to the effects from varying $\Omega_\Lambda$, except that the latter also changes the peak heights due to the late-time integrated Sachs-Wolfe (ISW) effect~\cite{Dodelson-book}. 

\subsection{Constraints on $\lambda_D$ and $z_T$}

%
\begin{figure}[th!]
\begin{center}
\hspace*{-0.0cm}
\includegraphics[width=0.48\textwidth, clip=true,viewport= 0 0 595 400]{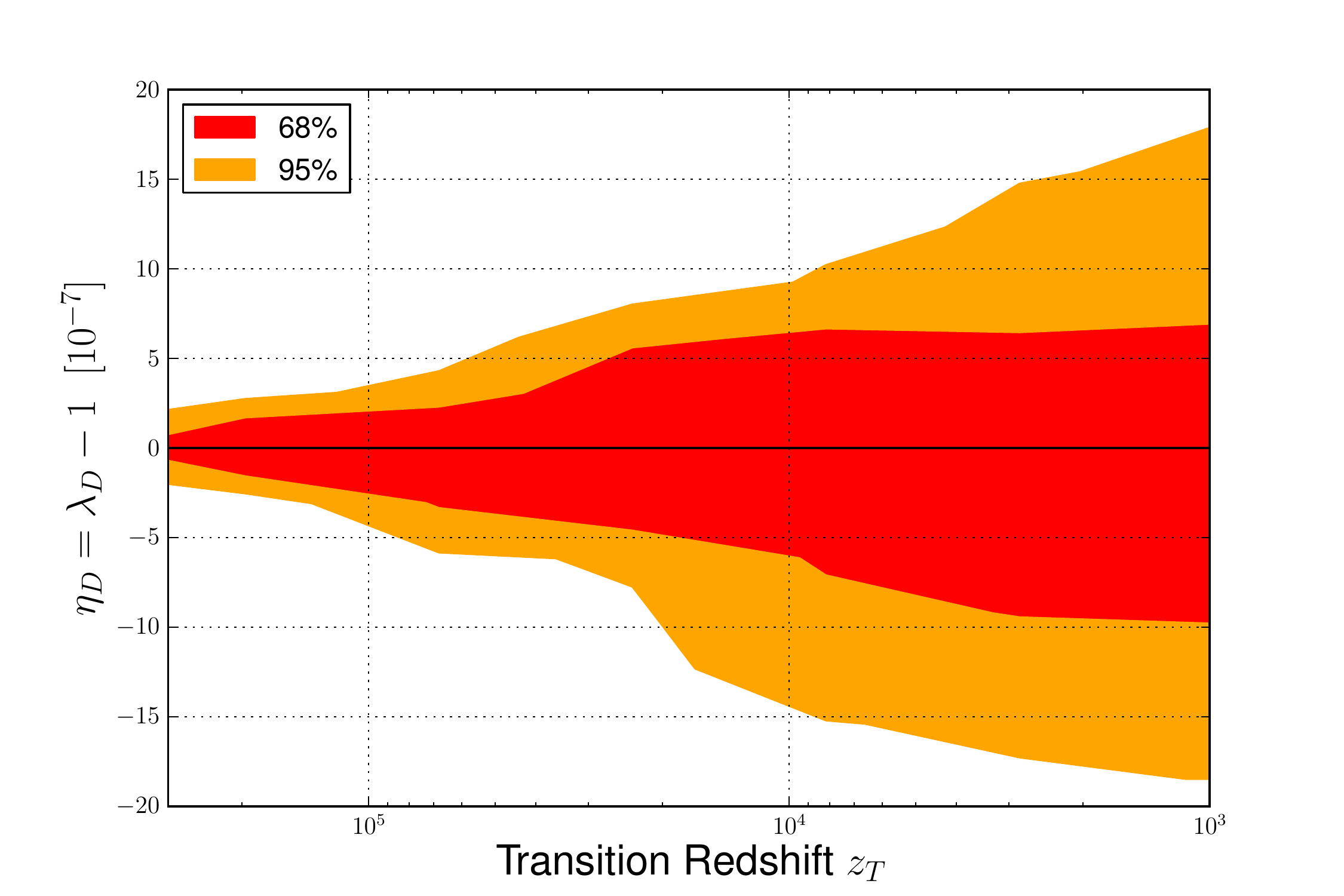} \hspace{2mm}
\includegraphics[width=0.48\textwidth, clip=true,viewport= 0 0 595 400]{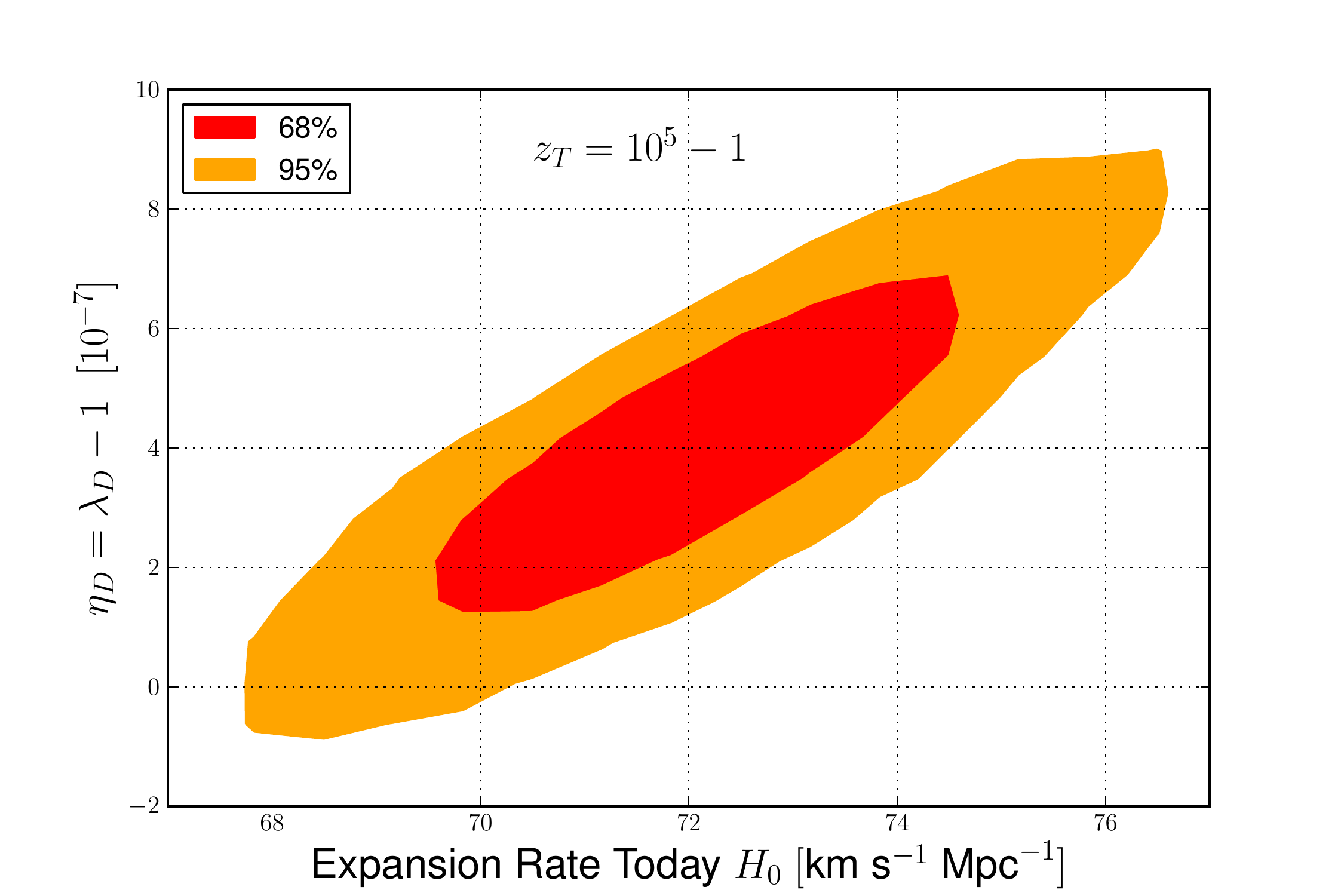}
\caption{{\bf Left panel:} 68\% and 95\% likelihood contours in the $\eta_{D} = \lambda_{D}-1$ and $z_{T}$ plane.
{\bf Right panel:} 68\% and 95\% likelihood contours in the $\eta_{D} = \lambda_{D}-1$ and $H_{0}$ plane. Here $z_{T} =  10^{5}-1$ is fixed to be a large value to provide a stringent constraint. Both plots use the dataset of Planck+Lensing+HST.
}
\label{fig:LvsZconst}
\end{center}
\end{figure}
Fig. \ref{fig:LvsZconst} shows the constraint on the parameter $\eta_{D} = \lambda_{D}-1$. We see that the constraint on $\eta_{D}$ becomes much stronger when the fluids decouple at  high redshift ({\it i.e.}, an earlier time). This is expected because the Friedmann equation is very sensitive to changes in the initial conditions. At high redshift, if $\lambda_{D}$ is not extremely close to one, an extraordinary amount of fine tuning is required to produce the observed universe. In the right panel, where we have fixed $z_{T} =  10^{5}-1$, we show the two-dimensional contour in $\eta_D$ and the current expansion rate $H_0$. The Planck three year measurement of $H_0=(67.3\pm1.3)~\mbox{km}\,\mbox{s}^{-1}\,\mbox{Mpc}^{-1}$~\cite{Ade:2013kta}, has some tension with the HST measurement of $H_0=(74.8\pm3.1)~\mbox{km}\,\mbox{s}^{-1}\,\mbox{Mpc}^{-1}$~\cite{Riess:2011yx}. As one can see, by allowing a non-zero value of $\eta_D$, one can resolve the tension between these measurements in this dark matter WEP breaking framework.

\section{Discussion and Conclusions}
\label{sec:conclusion}

In this paper we have updated the constraint on Newton's gravitational constant $G$ for all matter using the latest available cosmological data. We found a tension with the CODATA standard value for the gravitational constant which is independent of the specific combination of experimental data used in our analysis. The most stringent cosmological constraint on $G$ comes from the combined dataset of Planck+ACT/SPT+Lensing+BAO+HST+BBN, and this result has a tension with the CODATA value at a significance of 2.2$\sigma$. More insight on this discrepancy will come soon with the release of the 2015 Planck data~\cite{Ade:2015xua}. If the best fit for $G$ still tends high, this may be a hint for a new physics in cosmology and the need to go beyond the $\Lambda\mbox{CDM}$ standard cosmological model.

We also considered a traditional fifth force model to mimic the effect of equivalence principle breaking in the dark matter sector. In this type of model, the interaction between two dark matter particles is modified by the presence of an ultra light scalar, which mediates a long range attractive force. The main observable effect of this fifth force is to slow the decay of the gravitational potential at late times due to the expansion of the universe, adding power at low-$\ell$ through the Integrated Sachs-Wolfe effect. Using the 2013 Planck data to constrain this effect, we found that the coupling strength of this fifth force relative to the gravitational force must be less than $10^{-4}$. This constraint is independent of the mediator mass for the region of parameter space which does not modify the cosmological background evolution.

Finally, we considered true WEP principle breaking in the dark matter sector ({\it i.e.}, an explicit difference between dark matter gravitational and inertial mass). General relativity, being a geometrical theory, encodes this equivalence by construction. As such, to break the WEP in the dark matter sector we follow a phenomenological approach, motivating the modifications to the cosmological equations from post-Newtonian and non-relativistic fluid dynamics where this principle can be easily broken. Providing a fully consistent theory that encodes this phenomenology is beyond the scope of this work. The main thing to be learned here is that such modifications drastically modify the evolution of the cosmological background. The observable effect, even for extremely small degrees of WEP breaking, is to cause a visible, dark energy like shift the location of the peaks in the CMB power spectrum. Within the framework of this phenomenological model, we found that the ratio of the dark matter gravitational to inertial mass is constrained to be unity at the $10^{-6}$ level when the two fluids decouple at times earlier than approximately the recombination time.

\vspace{3mm}
\acknowledgments
We thank Amol Upadhye for useful discussion. 
The work is supported by the U. S. Department of Energy under the contract DE-FG-02-95ER40896 and by the National Science Foundation under grants OPP-0236449 and PHY-0236449.

\appendix
\section{Linear Perturbation Equations for the Two-Fluid Description}
\label{app:linear-equations}
For our two-fluid description, we have two independent background evolution functions: $a(t)$ for ordinary baryons and $a_D(t)$ for dark matter. We use $a(t)$ as a clock such that the linear perturbation for ordinary baryons will be unchanged, except that the common gravitational potential is also sourced by dark matter. 

Following the Newtonian approach in Mukhanov~\cite{Mukhanov-book}, the basic equations that govern the baryon and dark matter perturbations are:
\beqa
&&\frac{\partial\rho}{\partial t}+\nabla_x(\rho\,\textbf{v})=0 \,, \\
&& \frac{\partial\textbf{v}}{\partial t}+(\textbf{v}\cdot\nabla_x)\textbf{v}  +\nabla_x \phi =0 \,, \\
&&\nabla_x^{2}\phi= 4\pi G_N \rho  \,, 
\eeqa
where are continuity, Euler and Poisson equations, respectively. For the dark matter field, one can separate the fields into the zeroth and leading order as
\beqa
\rho_D = \rho_D^0 + \delta \rho_D \,, \qquad 
\textbf{v}_D = \textbf{v}_D^0 + \delta \textbf{v}_D \,, \qquad
\phi = \phi^0 + \delta\phi \,.
\eeqa
Substituting the perturbations into the above three equations, we obtain the following equations
\beqa
&&\frac{\partial\delta\rho_{D}}{\partial t}+\rho_{D}^{0}\nabla\cdot\delta\textbf{v}_{D}+\nabla_x(\delta\rho_{D} \cdot \textbf{v}_{D}^{0}) =0  \,, \\
&&\frac{\partial\delta\textbf{v}_{D}}{\partial t} + (\textbf{v}_{D}^{0}\cdot\nabla_x)\delta\textbf{v}_{D}
+(\delta\textbf{v}_{D}\cdot\nabla_x)\textbf{v}_{D}^{0} +\nabla_x\delta\phi=0 \,,  \\
&&\nabla_x^{2}\delta\phi= 4\pi G_N\, \left[\delta\rho_{b}+\lambda_D\,\delta\rho_{D} \right] \,,
\eeqa
The last Poisson equation should be modified when including radiation and dark energy densities. 

For the continuity equation and using the background field $\textbf{v}_D^0=H_{D}\,\textbf{x}$, we have $3H_{D}=\nabla_{x}\cdot\textbf{v}_{D}^0$. Here, the Hubble parameter is defined as $H_D = \frac{1}{a_D}\frac{da_D}{dt}$. The zeroth order continuity equation has the canonical relation $d{\rho}_{D}^0/dt = -3H_{D}\rho_{D}^0$. Defining the dimensionless quantity, $\delta_D = \delta \rho_D / \rho_D^0$, we have the first-order continuity equation as 
\beqa
\frac{\partial\delta_{D}}{\partial t}-(\nabla_{x}\cdot\textbf{v}_{D}^0) \ts \delta_{D}+\nabla_x\cdot\delta\textbf{v}_{D}+\nabla_x(\delta_{D} \cdot \textbf{v}_{D}^{0}) =0 \,.
\eeqa

In order to only use the ordinary baryon scale factor as a measure for time, we transform the dark matter equations to the \emph{baryon} co-moving frame with $\textbf{x} = a(t)\textbf{q}$.  The relation between the time-derivative's in the universe and co-moving frames is
\beqa
\left( \frac{\partial}{\partial t} \right)_{x}=\left( \frac{\partial}{\partial t} \right)_{q}-(\textbf{v}_{b}^0\cdot\nabla_{x}) = \left( \frac{\partial}{\partial t} \right)_{q}-\left(\frac{H}{H_D}\textbf{v}_{D}^0\cdot\nabla_{x}\right)  \,, \qquad 
\nabla_{x}=\frac{1}{a}\nabla_{q} \equiv \frac{1}{a} \nabla \,,
\eeqa
where we have used the relations $\textbf{v}_b^0 = H\, \textbf{x}$ and $\textbf{v}_D^0 = H_D\, \textbf{x}$ to have $\textbf{v}_b^0 = H \textbf{v}_D^0/H_D$. So, in the  \emph{baryon} co-moving frame, we have
\beqa
&&\frac{\partial\delta_{D}}{\partial t}+\frac{1}{a}\left(1-\frac{H}{H_{D}}\right)(\textbf{v}_{D}^0\cdot\nabla)\delta_{D}+\frac{1}{a}\nabla\cdot\delta\textbf{v}_{D} =0 \,, \\
&&\frac{\partial\delta\textbf{v}_{D}}{\partial t} + \frac{1}{a}\left(1-\frac{H}{H_{D}}\right)(\textbf{v}_{D}^{0}\cdot\nabla)\delta\textbf{v}_{D}
+H_{D}\ts \delta\textbf{v}_{D} +\frac{1}{a}\nabla\delta\phi=0 \,, 
\label{eq:eulerEQ}  \\
&&\nabla^{2}\delta\phi= 4\pi G_N\,a^{2} \left[\rho_{b}^{0}\,\delta_{b}+\lambda_D\,\rho_{D}^{0}\,\delta_{D} \right] \,.
\eeqa
Defining the divergence of the velocity field as $\theta_D = \nabla \delta\textbf{v}_D$ and changing the ordinary-time derivative to conformal-time derivative, we have  
\beqa
&&\dot{\delta}_{D}+\left(1-\frac{{\cal H}}{{\cal H}_{D}}\right)(\textbf{v}_{D}^0\cdot\nabla)\delta_{D}+\theta_{D} =0    \,, \\
&& \dot{\theta}_{D} + \left(1-\frac{{\cal H}}{{\cal H}_{D}}\right)\nabla\cdot[(\textbf{v}_{D}^{0}\cdot\nabla)\delta\textbf{v}_{D}]
+{\cal H}_{D}\theta_{D} +\nabla^{2}\delta\phi=0 \,, \\
&&\nabla^{2}\delta\phi= 4\pi G_N\,a^{2} \left[\rho_{b}^{0}\,\delta_{b}+\lambda_D\,\rho_{D}^{0}\,\delta_{D} \right]\,.
\eeqa
Here, we have used ${\cal H} = a H$ and ${\cal H}_D = a H_D$. After some algebra, one has 
\beqa
\nabla\cdot[(\textbf{v}_{D}^{0}\cdot\nabla)\delta\textbf{v}_{D}] &=& 3{\cal H}_{D}\theta_{D} + \textbf{v}_{D}^{0}\cdot\nabla^{2}\delta\textbf{v}_{D}+(\textbf{v}_{D}^{0}\cdot\nabla)\theta_{D}  \\
&=&  3{\cal H}_{D}\theta_{D} + 2(\textbf{v}_{D}^{0}\cdot\nabla)\theta_{D} \,.
\eeqa
In the second line of the above equation, we have used $\nabla^{2}\textbf{A} = \nabla(\nabla\cdot\textbf{A})-\nabla\times(\nabla\times\textbf{A})$ and assumed curl-less velocity perturbations ({\it i.e.} $\nabla\times\delta\textbf{v}_{D}=0$). We now define the operator $\hat{\mathcal{D}}$
\beq
\hat{\mathcal{D}} = \left(1-\frac{{\cal H}}{{\cal H}_{D}}\right)(\textbf{v}_{D}^{0}\cdot\nabla) \,. 
\eeq
The continuity and Euler equations become
\beqa
&& \dot{\delta}_{D}+\hat{\mathcal{D}}\delta_{D}+\theta_{D} =0  \,, \\
&&  \dot{\theta}_{D} +(4{\cal H}_{D}-3{\cal H}+ 2\hat{\mathcal{D}})\theta_{D} +\nabla^{2}\delta\phi=0 \,.
\eeqa
When ${\cal H}_{D}={\cal H}$ ({\it i.e.} normal cosmology), the operator $\hat{\mathcal{D}}$ is exactly zero and the equations reduce to the equation in the standard cosmology~\cite{Ma:1995ey}. The appearance of this new operator indicates the breaking of  equivalence principle. Since we chose the frame to be the baryon co-moving frame, the operator $\hat{\mathcal{D}}$ appears as an artifact in the dark matter
equations to convert from the dark matter co-moving frame to the baryon co-moving frame. Using the Fourier transformation in Appendix~\ref{app:fourier-transoformation}, we have the corresponding equations in momentum space as
\beqa
&&\dot{\delta}_{D}\,+\,({\cal H}- {\cal H}_D)\,( 3 + k \ts \partial_{k})\,\delta_{D} \,=\,3\,\delta\dot{\phi}-\theta_{D} \,, \label{eq:delta_D-final} \\
&&\dot{\theta}_{D}\,+\,{\cal H}\,\theta_D \,+\,2\,({\cal H}- {\cal H}_D)\,( 1 + k \ts \partial_{k} )\,\theta_{D} \,=\,  k^{2}\,\delta\psi\,,  \\
&&k^2\,\delta\phi + 3 {\cal H} ( \delta \dot{\phi} + {\cal H} \delta\psi)= - 4\pi G_N\,a^{2} \left[\rho_{b}^{0}\,\delta_{b}+\lambda_D\,\rho_{D}^{0}\,\delta_{D} \right]  \,.
\label{eq:delta_phi-final}
\eeqa
In the Newtonian gauge and with the assumption of no shear, we have $\delta \phi = \delta \psi$. In the right-hand side of the first equation of Eq.~(\ref{eq:delta_D-final}) and in the left-hand side of the third equation of Eq.~(\ref{eq:delta_phi-final}), we have also added an extra term to account for the geometry part from general relativity, which is beyond this post-Newtonian description.

\section{Fourier Transformation of the New Operator $\hat{\mathcal{D}}$}
\label{app:fourier-transoformation}
Let us consider the Fourier transform of this new operator acting on some function $f(\textbf{x})=f(a\cdot\textbf{q})$
\beqa
\mathcal{F}(\hat{\mathcal{D}}f) = \int_{V} d^{3}q \ts e^{-i\textbf{k}\cdot\textbf{q}}\hat{\mathcal{D}}f 
=\left(1-\frac{{\cal H}}{{\cal H}_{D}}\right) \int_{V} d^{3}q \ts e^{-i\textbf{k}\cdot\textbf{q}}(\textbf{v}_{D}^{0}\cdot\nabla f) \,.
\eeqa
Integrating it by parts, one obtains
\beqa
\mathcal{F}(\hat{\mathcal{D}}f) =\left(1-\frac{{\cal H}}{{\cal H}_{D}}\right) \left[ \int_{\partial V} 
\ts e^{-i\textbf{k}\cdot\textbf{q}}(\textbf{v}_{D}^{0}\cdot \hat{\textbf{n}}) f \ts dS - 
\int_{V} d^{3}q  \ts f \ts \nabla\cdot(e^{-i\textbf{k}\cdot\textbf{q}}\ts\textbf{v}_{D}^{0})\right] \,, 
\eeqa
where the surface integral vanishes when integrating over all space as long as $f$ decays fast at infinity. So we have
\beqa
\mathcal{F}(\hat{\mathcal{D}}f) &=& \left(\frac{{\cal H}}{{\cal H}_{D}}-1\right)
\int_{V} d^{3}q  \ts f e^{-i\textbf{k}\cdot\textbf{q}}\left[ -i(\textbf{k}\cdot\textbf{v}_{D}^{0}) + \nabla\cdot \textbf{v}_{D}^{0}\right]  \nonumber  \\
& =&\left(\frac{{\cal H}}{{\cal H}_{D}}-1\right)
\int_{V} d^{3}q  \ts f e^{-i\textbf{k}\cdot\textbf{q}}\left[ -i{\cal H}_{D}(\textbf{k}\cdot\textbf{q}) + 3{\cal H}_{D}\right]  \nonumber  \\
&=& \left({\cal H}-{\cal H}_{D}\right)\left[3\mathcal{F}(f)+\textbf{k}\cdot\nabla_{\textbf{k}}
\int_{V} d^{3}q  \ts f  \ts e^{-i\textbf{k}\cdot\textbf{q}} \right]  \nonumber \\
&=& \left({\cal H}-{\cal H}_{D}\right)\left[3+(\textbf{k}\cdot\nabla_{\textbf{k}})
\right]\mathcal{F}(f) \,.
\eeqa
Where we have used $\textbf{v}_D^0 = H_D\, \textbf{x} = {\cal H}_D \, \textbf{q}$.

\providecommand{\href}[2]{#2}\begingroup\raggedright\endgroup


\begin{thebibliography}{10}

\bibitem{Adelberger:2003zx}
E.~Adelberger, B.~R. Heckel, and A.~Nelson, {\it {Tests of the gravitational
  inverse square law}},  {\em Ann.Rev.Nucl.Part.Sci.} {\bf 53} (2003) 77--121,
  [\href{http://arxiv.org/abs/hep-ph/0307284}{{\tt hep-ph/0307284}}].

\bibitem{Mohr:2012tt}
P.~J. Mohr, B.~N. Taylor, and D.~B. Newell, {\it {CODATA Recommended Values of
  the Fundamental Physical Constants: 2010}},  {\em Rev.Mod.Phys.} {\bf 84}
  (2012) 1527--1605, [\href{http://arxiv.org/abs/1203.5425}{{\tt
  arXiv:1203.5425}}].

\bibitem{cold-atom}
G.~Rosi, F.~Sorrentino, L.~Cacciapuoti, M.~Prevedelli, and G.~M. Tino, {\it
  {Precision measurement of the Newtonian gravitational constant using cold
  atoms}},  {\em Nature} {\bf 510} (2014), no.~7506, 518--521.

\bibitem{Zahn:2002rr}
O.~Zahn and M.~Zaldarriaga, {\it {Probing the Friedmann equation during
  recombination with future CMB experiments}},  {\em Phys.Rev.} {\bf D67}
  (2003) 063002, [\href{http://arxiv.org/abs/astro-ph/0212360}{{\tt
  astro-ph/0212360}}].

\bibitem{Umezu:2005ee}
K.-i. Umezu, K.~Ichiki, and M.~Yahiro, {\it {Cosmological constraints on
  Newton's constant}},  {\em Phys.Rev.} {\bf D72} (2005) 044010,
  [\href{http://arxiv.org/abs/astro-ph/0503578}{{\tt astro-ph/0503578}}].

\bibitem{Galli:2009pr}
S.~Galli, A.~Melchiorri, G.~F. Smoot, and O.~Zahn, {\it {From Cavendish to
  PLANCK: Constraining Newton's Gravitational Constant with CMB Temperature and
  Polarization Anisotropy}},  {\em Phys.Rev.} {\bf D80} (2009) 023508,
  [\href{http://arxiv.org/abs/0905.1808}{{\tt arXiv:0905.1808}}].

\bibitem{Ade:2013zuv}
{\bf Planck} Collaboration, P.~Ade et~al., {\it {Planck 2013 results. XVI.
  Cosmological parameters}},  {\em Astron.Astrophys.} {\bf 571} (2014) A16,
  [\href{http://arxiv.org/abs/1303.5076}{{\tt arXiv:1303.5076}}].

\bibitem{Ade:2015xua}
{\bf Planck} Collaboration, P.~Ade et~al., {\it {Planck 2015 results. XIII.
  Cosmological parameters}},  \href{http://arxiv.org/abs/1502.01589}{{\tt
  arXiv:1502.01589}}.

\bibitem{Sievers:2013ica}
{\bf Atacama Cosmology Telescope} Collaboration, J.~L. Sievers et~al., {\it
  {The Atacama Cosmology Telescope: Cosmological parameters from three seasons
  of data}},  {\em JCAP} {\bf 1310} (2013) 060,
  [\href{http://arxiv.org/abs/1301.0824}{{\tt arXiv:1301.0824}}].

\bibitem{Hou:2012xq}
Z.~Hou, C.~Reichardt, K.~Story, B.~Follin, R.~Keisler, et~al., {\it
  {Constraints on Cosmology from the Cosmic Microwave Background Power Spectrum
  of the 2500-square degree SPT-SZ Survey}},  {\em Astrophys.J.} {\bf 782}
  (2014) 74, [\href{http://arxiv.org/abs/1212.6267}{{\tt arXiv:1212.6267}}].

\bibitem{Wagner:2012}
T.~Wagner, S.~Schlamminger, J.~Gundlach, and E.~Adelberger, {\it
  {Torsion-balance tests of the weak equivalence principle}},
  \href{http://arxiv.org/abs/1207.2442}{{\tt arXiv:1207.2442}}.

\bibitem{Friedman:1991dj}
J.~A. Frieman and B.-A. Gradwohl, {\it {Dark matter and the equivalence
  principle}},  {\em Phys.Rev.Lett.} {\bf 67} (1991) 2926--2929.

\bibitem{Bean:2001ys}
R.~Bean, {\it {Perturbation evolution with a nonminimally coupled scalar
  field}},  {\em Phys.Rev.} {\bf D64} (2001) 123516,
  [\href{http://arxiv.org/abs/astro-ph/0104464}{{\tt astro-ph/0104464}}].

\bibitem{Gubser:2004uh}
S.~S. Gubser and P.~Peebles, {\it {Structure formation in a string inspired
  modification of the cold dark matter model}},  {\em Phys.Rev.} {\bf D70}
  (2004) 123510, [\href{http://arxiv.org/abs/hep-th/0402225}{{\tt
  hep-th/0402225}}].

\bibitem{Nusser:2004qu}
A.~Nusser, S.~S. Gubser, and P.~Peebles, {\it {Structure formation with a
  long-range scalar dark matter interaction}},  {\em Phys.Rev.} {\bf D71}
  (2005) 083505, [\href{http://arxiv.org/abs/astro-ph/0412586}{{\tt
  astro-ph/0412586}}].

\bibitem{Bean:2008ac}
R.~Bean, E.~E. Flanagan, I.~Laszlo, and M.~Trodden, {\it {Constraining
  Interactions in Cosmology's Dark Sector}},  {\em Phys.Rev.} {\bf D78} (2008)
  123514, [\href{http://arxiv.org/abs/0808.1105}{{\tt arXiv:0808.1105}}].

\bibitem{Mukhanov-book}
V.~Mukhanov, {\em {Physical Foundations of Cosmology}}.
\newblock Cambridge University Press, 2005.

\bibitem{Kesden:2006zb}
M.~Kesden and M.~Kamionkowski, {\it {Galilean Equivalence for Galactic Dark
  Matter}},  {\em Phys.Rev.Lett.} {\bf 97} (2006) 131303,
  [\href{http://arxiv.org/abs/astro-ph/0606566}{{\tt astro-ph/0606566}}].

\bibitem{Galli:2010it}
S.~Galli, M.~Martinelli, A.~Melchiorri, L.~Pagano, B.~D. Sherwin, et~al., {\it
  {Constraining Fundamental Physics with Future CMB Experiments}},  {\em
  Phys.Rev.} {\bf D82} (2010) 123504,
  [\href{http://arxiv.org/abs/1005.3808}{{\tt arXiv:1005.3808}}].

\bibitem{Ade:2014lua}
{\bf Planck} Collaboration, P.~Ade et~al., {\it {Planck intermediate results.
  XXIV. Constraints on variation of fundamental constants}},
  \href{http://arxiv.org/abs/1406.7482}{{\tt arXiv:1406.7482}}.

\bibitem{Ma:1995ey}
C.-P. Ma and E.~Bertschinger, {\it {Cosmological perturbation theory in the
  synchronous and conformal Newtonian gauges}},  {\em Astrophys.J.} {\bf 455}
  (1995) 7--25, [\href{http://arxiv.org/abs/astro-ph/9506072}{{\tt
  astro-ph/9506072}}].

\bibitem{Seljak:1996is}
U.~Seljak and M.~Zaldarriaga, {\it {A Line of sight integration approach to
  cosmic microwave background anisotropies}},  {\em Astrophys.J.} {\bf 469}
  (1996) 437--444, [\href{http://arxiv.org/abs/astro-ph/9603033}{{\tt
  astro-ph/9603033}}].

\bibitem{Audren:2012wb}
B.~Audren, J.~Lesgourgues, K.~Benabed, and S.~Prunet, {\it {Conservative
  Constraints on Early Cosmology: an illustration of the Monte Python
  cosmological parameter inference code}},  {\em JCAP} {\bf 1302} (2013) 001,
  [\href{http://arxiv.org/abs/1210.7183}{{\tt arXiv:1210.7183}}].

\bibitem{Lesgourgues:2011re}
J.~Lesgourgues, {\it {The Cosmic Linear Anisotropy Solving System (CLASS) I:
  Overview}},  \href{http://arxiv.org/abs/1104.2932}{{\tt arXiv:1104.2932}}.

\bibitem{Blas:2011rf}
D.~Blas, J.~Lesgourgues, and T.~Tram, {\it {The Cosmic Linear Anisotropy
  Solving System (CLASS) II: Approximation schemes}},  {\em JCAP} {\bf 1107}
  (2011) 034, [\href{http://arxiv.org/abs/1104.2933}{{\tt arXiv:1104.2933}}].

\bibitem{Gelman:1992zz}
A.~Gelman and D.~B. Rubin, {\it {Inference from Iterative Simulation Using
  Multiple Sequences}},  {\em Statist.Sci.} {\bf 7} (1992) 457--472.

\bibitem{Ade:2013kta}
{\bf Planck} Collaboration, P.~Ade et~al., {\it {Planck 2013 results. XV. CMB
  power spectra and likelihood}},  \href{http://arxiv.org/abs/1303.5075}{{\tt
  arXiv:1303.5075}}.

\bibitem{Riess:2011yx}
A.~G. Riess, L.~Macri, S.~Casertano, H.~Lampeitl, H.~C. Ferguson, et~al., {\it
  {A 3
  Telescope and Wide Field Camera 3}},  {\em Astrophys.J.} {\bf 730} (2011)
  119, [\href{http://arxiv.org/abs/1103.2976}{{\tt arXiv:1103.2976}}].

\bibitem{SDSS7}
{\bf SDSS} Collaboration, W.~J. Percival et~al., {\it {Baryon Acoustic
  Oscillations in the Sloan Digital Sky Survey Data Release 7 Galaxy Sample}},
  {\em Mon.Not.Roy.Astron.Soc.} {\bf 401} (2010) 2148--2168,
  [\href{http://arxiv.org/abs/0907.1660}{{\tt arXiv:0907.1660}}].

\bibitem{SDSS9}
L.~Anderson, E.~Aubourg, S.~Bailey, F.~Beutler, A.~S. Bolton, et~al., {\it {The
  clustering of galaxies in the SDSS-III Baryon Oscillation Spectroscopic
  Survey: Measuring $D_A$ and $H$ at $z=0.57$ from the Baryon Acoustic Peak in
  the Data Release 9 Spectroscopic Galaxy Sample}},
  \href{http://arxiv.org/abs/1303.4666}{{\tt arXiv:1303.4666}}.

\bibitem{Beutler:2011hx}
F.~Beutler, C.~Blake, M.~Colless, D.~H. Jones, L.~Staveley-Smith, et~al., {\it
  {The 6dF Galaxy Survey: Baryon Acoustic Oscillations and the Local Hubble
  Constant}},  {\em Mon.Not.Roy.Astron.Soc.} {\bf 416} (2011) 3017--3032,
  [\href{http://arxiv.org/abs/1106.3366}{{\tt arXiv:1106.3366}}].

\bibitem{Hill:1987bm}
C.~T. Hill and G.~G. Ross, {\it {PseudoGoldstone Bosons and New Macroscopic
  Forces}},  {\em Phys.Lett.} {\bf B203} (1988) 125.

\bibitem{Koivisto:2005nr}
T.~Koivisto, {\it {Growth of perturbations in dark matter coupled with
  quintessence}},  {\em Phys.Rev.} {\bf D72} (2005) 043516,
  [\href{http://arxiv.org/abs/astro-ph/0504571}{{\tt astro-ph/0504571}}].

\bibitem{Morris:2013hua}
S.~C.~F. Morris, A.~M. Green, A.~Padilla, and E.~R.~M. Tarrant, {\it
  {Cosmological effects of coupled dark matter}},  {\em Phys.Rev.} {\bf D88}
  (2013), no.~8 083522, [\href{http://arxiv.org/abs/1304.2196}{{\tt
  arXiv:1304.2196}}].

\bibitem{Nolta:2008ih}
{\bf WMAP} Collaboration, M.~Nolta et~al., {\it {Five-Year Wilkinson Microwave
  Anisotropy Probe (WMAP) Observations: Angular Power Spectra}},  {\em
  Astrophys.J.Suppl.} {\bf 180} (2009) 296--305,
  [\href{http://arxiv.org/abs/0803.0593}{{\tt arXiv:0803.0593}}].

\bibitem{Adelberger:2006dh}
E.~Adelberger, B.~R. Heckel, S.~A. Hoedl, C.~Hoyle, D.~Kapner, et~al., {\it
  {Particle Physics Implications of a Recent Test of the Gravitational Inverse
  Sqaure Law}},  {\em Phys.Rev.Lett.} {\bf 98} (2007) 131104,
  [\href{http://arxiv.org/abs/hep-ph/0611223}{{\tt hep-ph/0611223}}].

\bibitem{Dodelson-book}
S.~Dodelson, {\em {Modern Cosmology}}.
\newblock Academic Press, 2003.

\end{thebibliography}
\end{document}